\documentclass[sn-mathphys,Numbered]{sn-jnl}

\usepackage{graphicx}
\usepackage{multirow}
\usepackage{amsmath,amssymb,amsfonts,amsbsy}
\usepackage{amsthm}
\usepackage{mathrsfs}
\usepackage[title]{appendix}
\usepackage{xcolor}
\usepackage{textcomp}
\usepackage{manyfoot}
\usepackage{booktabs}
\usepackage{algorithm}
\usepackage{algorithmicx}
\usepackage{algpseudocode}
\usepackage{listings}
\usepackage{mathtools}
\usepackage{lmodern}
\usepackage{subfiles}
\usepackage{array}
\usepackage{bm}

\begin{document}

\title{Scenario-driven optimization of passive vehicle suspensions: explaining the effectiveness of asymmetric damping}

\author*[1]{\fnm{José Geraldo} \sur{Telles Ribeiro}}\email{jose.ribeiro@uerj.br}

\author[2,3]{\fnm{Americo} \sur{Cunha Jr}}\email{americo@lncc.br}

\affil[1]{\orgdiv{Department of Mechanical Engineering}, \orgname{Rio de Janeiro State University}, \orgaddress{\street{Rua  São Francisco Xavier, 524}, \city{Rio de Janeiro}, \postcode{20550-900}, \state{RJ}, \country{Brazil}}}

\affil[2]{\orgdiv{Department of Applied Mathematics}, \orgname{Rio de Janeiro State University}, \orgaddress{\street{Rua  São Francisco Xavier, 524}, \city{Rio de Janeiro}, \postcode{20550-900}, \state{RJ}, \country{Brazil}}}

\affil[3]{\orgname{National Laboratory of Scientific Computing}, \orgaddress{\street{Av. Getúlio Vargas, 333}, \city{Petrópolis}, \postcode{25651-075}, \state{RJ}, \country{Brazil}}}

\abstract{Asymmetric damping is widely used in passive vehicle suspensions, with rebound damping often recommended to exceed compression damping by a factor of two to three. Despite its prevalence, this guideline remains largely empirical and lacks a systematic derivation based on vehicle dynamics and excitation conditions. This paper presents a scenario-driven optimization framework that provides a principled explanation for the effectiveness of asymmetric damping. A minimal quarter-car model is employed to isolate the key mechanisms governing the trade-off between ride comfort, road holding, and transient response, using standardized ISO~8608 road excitations. Rebound and compression damping ratios are treated as independent design variables, and optimal configurations are identified via a stochastic Cross-Entropy algorithm applied to a non-convex, simulation-based objective function. Performance is assessed through ISO~2631 weighted RMS acceleration, tire--ground contact force variability, and settling time. The results show that symmetric damping is often sufficient under moderate excitation, whereas asymmetric damping becomes necessary under severe conditions, with commonly cited rebound-to-compression ratios emerging as scenario-dependent near-optimal solutions rather than universal constants.
}

\keywords{passive vehicle suspension; asymmetric damping; ride comfort; transient response; cross-entropy optimization}

\maketitle

\section{Introduction} 
\label{sec1}

Vehicles typically operate across a wide range of road surfaces that differ markedly in geometry, roughness, and material composition. Although automotive suspensions are designed according to common engineering principles, road excitations are highly heterogeneous, imposing distinct dynamic demands on the suspension system \cite{lee2025vehicle}. The interaction between vehicle and terrain governs not only ride comfort but also fundamental safety-related aspects such as tire–road adhesion, braking effectiveness, and directional stability \cite{Wong2008TheoryGroundVehicles,Rajamani2012VehicleDynamics,uys2007suspension}.

At the same time, vehicles themselves exhibit substantial variability in mass, load distribution, and functional purpose, ranging from lightweight passenger cars to heavy-duty and mid-heavy vehicles. Each category operates within a specific envelope of road conditions and velocity ranges, requiring suspension characteristics tailored to its typical use \cite{Rajamani2012VehicleDynamics,Gillespie1992Fundamentals,Bui2025SISS_HMDV_JVET,Zhang2025QZS_AirSuspension_JVET,Wong2008TheoryGroundVehicles}. A suspension system that is excessively compliant may improve comfort at low excitation levels but degrade handling and stability, whereas overly stiff configurations enhance control at the expense of ride comfort. Achieving an appropriate balance between comfort, stability, and safety therefore remains a central challenge in suspension design \cite{Gillespie1992Fundamentals,Rajamani2012VehicleDynamics,Dixon2017ShockAbsorberHandbook,Torun2025RailComfortDerailmentTradeoffJVETech}.

\begin{figure}[h!]
\centering
\includegraphics[width=130mm]{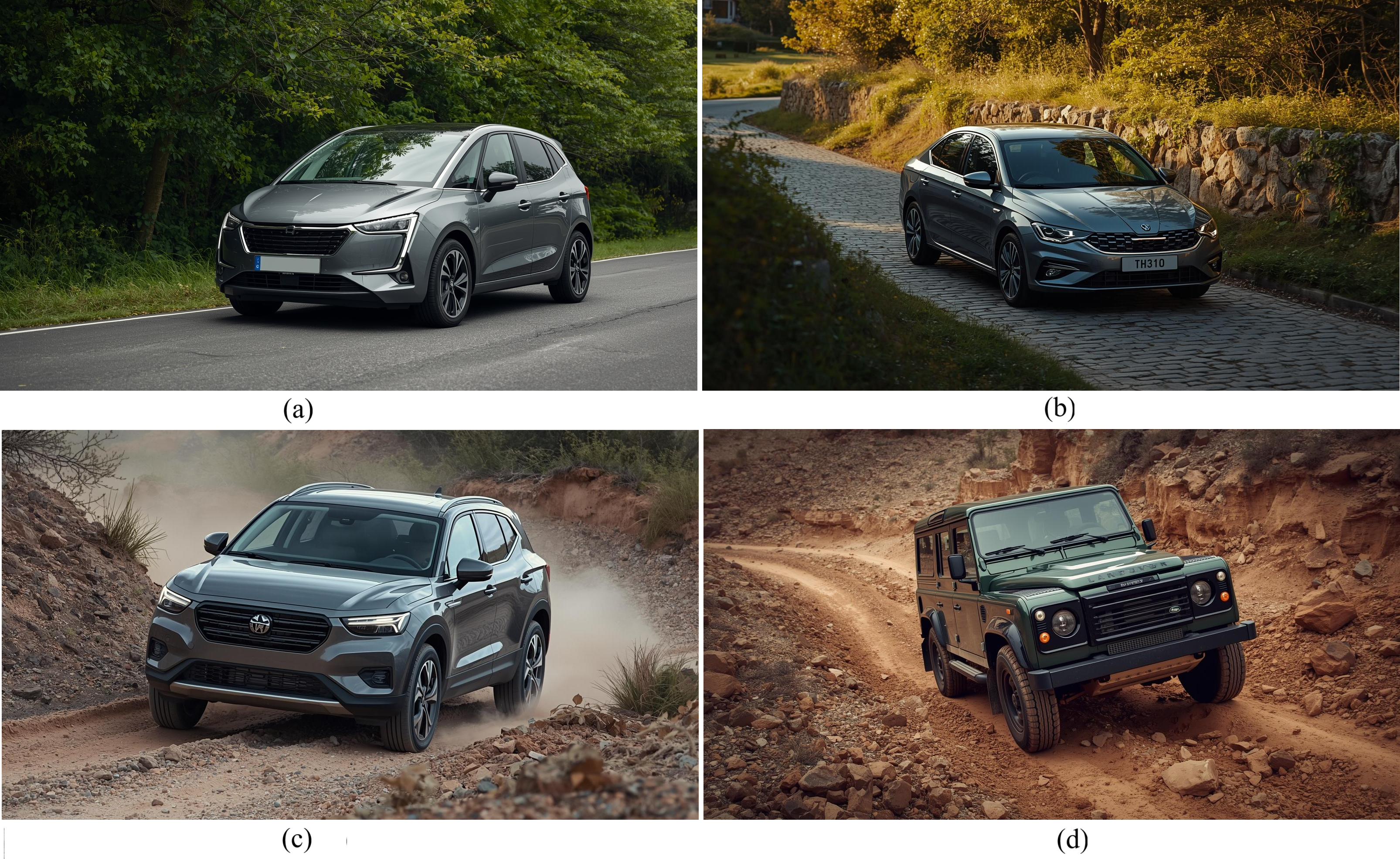}
\caption {Representative vehicle–road interaction scenarios illustrating the influence of surface condition and vehicle class on suspension response: (a) light vehicle operating on a well-paved asphalt surface, characterized by low-amplitude, high-frequency excitations and favorable adhesion; (b) light vehicle on a cobblestone road, where increased surface irregularity introduces higher excitation levels and degrades ride comfort and stability; (c) light vehicle operating off-road on an unpaved surface, experiencing large suspension displacements, elevated dynamic loads, and intermittent reductions in tire–road contact; and (d) mid-heavy vehicle operating off-road on an unpaved surface, where increased mass, long-travel suspension, and reinforced components improve adaptability and stability under severe terrain excitation. The images are illustrative conceptual representations (not experimental photographs) generated with an AI-based image-generation tool.}
\label{Figure_01}
\end{figure}

Figure \ref{Figure_01}(a)–(d) illustrates the diversity of vehicle–road interaction scenarios considered in this context. Figure \ref{Figure_01}(a) depicts a light vehicle traveling on a well-paved asphalt surface, where small-amplitude, high-frequency irregularities dominate the excitation. In contrast, Fig. \ref{Figure_01}(b) shows the same vehicle operating on a cobblestone road, characterized by pronounced surface irregularity and increased excitation severity. Figure \ref{Figure_01}(c) presents a light vehicle under off-road conditions, where large suspension displacements and intermittent loss of tire contact may occur. Finally, Fig. \ref{Figure_01}(d) illustrates a heavy vehicle operating off-road, highlighting how increased mass and suspension travel modify the dynamic response and improve terrain adaptability.

These examples emphasize that vehicle dynamics cannot be analyzed independently of the terrain. Road profile, vehicle mass, and suspension properties form a strongly coupled system whose dynamic behavior directly influences both comfort and safety \cite{lukosevicius2021investigation}. In addition, vehicle speed plays a critical role in this interaction. The same road irregularity can produce markedly different responses at different velocities. While smooth surfaces generally allow higher speeds without excessive discomfort, rough terrain amplifies suspension deflections, wheel accelerations, and the risk of tire detachment as speed increases. These effects are particularly severe for light vehicles, whose limited suspension travel and lower inertia make them sensitive to high-frequency excitations, whereas heavy vehicles experience larger dynamic loads that may threaten structural integrity or control when operating beyond safe velocity limits.

Over the past decades, numerous strategies have been proposed to mitigate vehicle vibrations and improve both ride comfort and road-holding performance, as in \cite{gonccalves2003optimization,goncalves2005road} that proposes using Flexible Multibody Dynamics. In the field of active suspensions, several control strategies have been investigated to enhance vehicle dynamics. Examples include fractional-order terminal sliding mode controllers \cite{rajendiran2020performance}, linear quadratic Gaussian (LQG) controllers \cite{pang2017design}, robust control approaches for electromagnetic active suspension systems \cite{van2013robust}, active anti-roll bar systems \cite{cronje2010improving}, model-free finite-time tracking control method \cite{wang2019implementation}, fuzzy controller for or uncertain vehicle suspension systems \cite{zhang2019reliable} and series active variable geometry suspension concepts \cite{arana2017series}.

Semi-active suspensions have also attracted considerable attention due to their ability to adapt damping characteristics while maintaining relatively low energy consumption \cite{soliman2019semi}. Many studies have explored magnetorheological dampers and related control strategies \cite{hemanth2016vertical,marathe2022development,yang2024phase,Narwade2022MRDamperRideComfortJVETech,verros2005design}. Other developments include inerter-based vibration isolation devices \cite{liu2022review,shen2026vehicle}, skyhook-based inerter configurations \cite{hu2017comfort}, sliding-mode and internal-model-based skyhook controllers \cite{jayabalan2018vibration, liu2019general}, and skyhook inertance constant-frequency control strategies \cite{zhang2026constant}. Hybrid semi-active vibration isolators combining nonlinear hysteretic models, such as the Bouc–Wen formulation, with viscoelastic elements have also been proposed to further improve suspension performance \cite{Kumar2025HybridSemiActiveHalfCarJVETech}.

Additional vibration-mitigation approaches have been investigated at different levels of the vehicle system. For example, seat-suspension systems have been developed to reduce the transmission of vibrations to vehicle occupants \cite{jamadar2021dynamic,fard2014effects,ning2016active}. Other studies have explored specialized vibration-control mechanisms, such as inerter-based nonlinear dissipation systems for mitigating vehicle shimmy \cite{Wang2025ShimmyInerterNonlinearJVETech} and passive suspension concepts employing negative-stiffness mechanisms to enhance vibration isolation \cite{Suman2021NegativeStiffnessSuspensionJVETech}.

Despite these advances, classical passive suspensions remain the most widely adopted solution in commercial vehicles due to their simplicity, robustness, and low energy requirements. In particular, passive configurations employing asymmetric damping have proven especially attractive for commercial vehicles and competition off-road vehicles, where reliability and mechanical simplicity are essential. Such systems exploit different compression and rebound damping characteristics to improve the trade-off between ride comfort and road-holding performance while maintaining the inherent advantages of passive suspension architectures.

Despite its widespread use in passive suspension design, asymmetric damping is most often introduced in the literature through empirical guidelines, commonly stating that rebound damping should exceed compression damping by a factor of two to three \cite{balike2010influence}. While these rules have proven effective in practice, they are typically stated without a clear mathematical justification or an explicit connection to the underlying excitation characteristics, vehicle properties, or competing performance objectives. As a consequence, damping asymmetry is frequently treated as a universal design prescription, rather than as the outcome of a structured decision process conditioned on operating scenarios and performance constraints.

The central objective of this work is to provide a principled, mathematically grounded explanation for why such empirical asymmetric damping rules emerge as effective design choices in practice, and to clarify the conditions under which they remain valid—or fail to do so. Rather than proposing new damper models or introducing additional sources of nonlinearity, the focus is deliberately placed on understanding how asymmetric damping arises naturally as an optimal solution of a constrained vibration mitigation problem when ride comfort, road holding, and transient response are simultaneously enforced. In this sense, the present study aims to formalize long-standing engineering intuition by embedding it within a reproducible optimization framework driven by realistic excitation scenarios.

To achieve this goal while preserving analytical transparency, a minimal yet representative modeling strategy is adopted. The quarter-car model is employed as a reduced-order description of vertical vehicle dynamics, retaining the essential interactions among the sprung mass, unsprung mass, suspension elements, and tire–road interface. Additional complexities—such as dry friction, detailed damper hysteresis, suspension geometry effects, or full-vehicle pitch and roll dynamics—are intentionally excluded. Likewise, semi-active suspension architectures and alternative vibration-control devices are outside the scope here, since the goal is to explain when passive asymmetric damping alone becomes structurally optimal \cite{Narwade2022MRDamperRideComfortJVETech,Kumar2025HybridSemiActiveHalfCarJVETech,Suman2021NegativeStiffnessSuspensionJVETech}. This choice is motivated by the need to isolate the fundamental mechanisms governing the comfort–stability trade-off and to avoid obscuring physical interpretation behind model-specific details. Within this simplified yet physically consistent framework, a scenario-driven\footnote{In this context, scenario-driven optimization refers to a design strategy in which suspension parameters are inferred from their performance across a set of representative operating scenarios, rather than being optimized for a single nominal condition. Each scenario is defined by a combination of road roughness, vehicle speed, mass, and excitation type, together with explicit comfort, stability, and transient performance constraints. The objective is not to identify a universally optimal configuration, but to determine which design structures consistently emerge as optimal under specific classes of operating conditions.} optimization approach demonstrates that empirical asymmetric damping ratios emerge as conditional optima whose effectiveness depends explicitly on road roughness, vehicle mass, speed, and transient performance requirements.

The rest of this paper is organized as follows: Section~\ref{sec2} introduces the quarter-car model with asymmetric damping and the road–excitation framework, and defines the simulation setup and operating conditions. Section~\ref{sec3} presents the performance metrics for ride comfort, road holding, and transient behavior, and formulates the scenario-driven non-convex optimization problem solved via the Cross-Entropy method. Section~\ref{sec4} reports and interprets the optimal suspension configurations across road classes, vehicle masses, and speeds, highlighting when and why damping asymmetry emerges as an optimal design feature. Finally, Section~\ref{sec5} summarizes the main conclusions and outlines directions for future work.
%
\section{Modeling the vehicle vertical dynamics} 
\label{sec2}

The objective of the present modeling strategy is not to reproduce the full complexity of real suspension systems, but to isolate the minimal set of dynamical mechanisms required to explain the effectiveness of empirical asymmetric damping rules. The model is therefore constructed as a controlled abstraction: it retains only those elements that directly influence the trade-off between ride comfort, road holding, and transient response, while deliberately excluding secondary effects that may obscure physical interpretation.

Following this modeling philosophy, the vertical dynamic behavior of a vehicle can be effectively studied using simplified representations that retain the essential physics while enabling analytical insight and computational efficiency. In this work, a quarter-car model \cite{hemanth2016vertical,silveira2017effects,verros2005design} is adopted to describe the interaction between the vehicle body, suspension elements, and tire dynamics under external excitation. This formulation provides the necessary foundation for analyzing ride comfort, road-holding capability, and load transfer mechanisms. To support subsequent simulations, the suspension and tire characteristics—including nonlinear and asymmetric behaviors—are explicitly defined, and representative road excitations are synthesized in accordance with the ISO 8608 standard \cite{ISO8608,agostinacchio2014vibrations}. Together, these components establish a consistent minimal framework for evaluating suspension performance across the different scenarios investigated in the results section.

\subsection{The quarter-car mechanical model}
\label{subsec21}

\begin{figure}
\centering
\includegraphics[width=110mm]{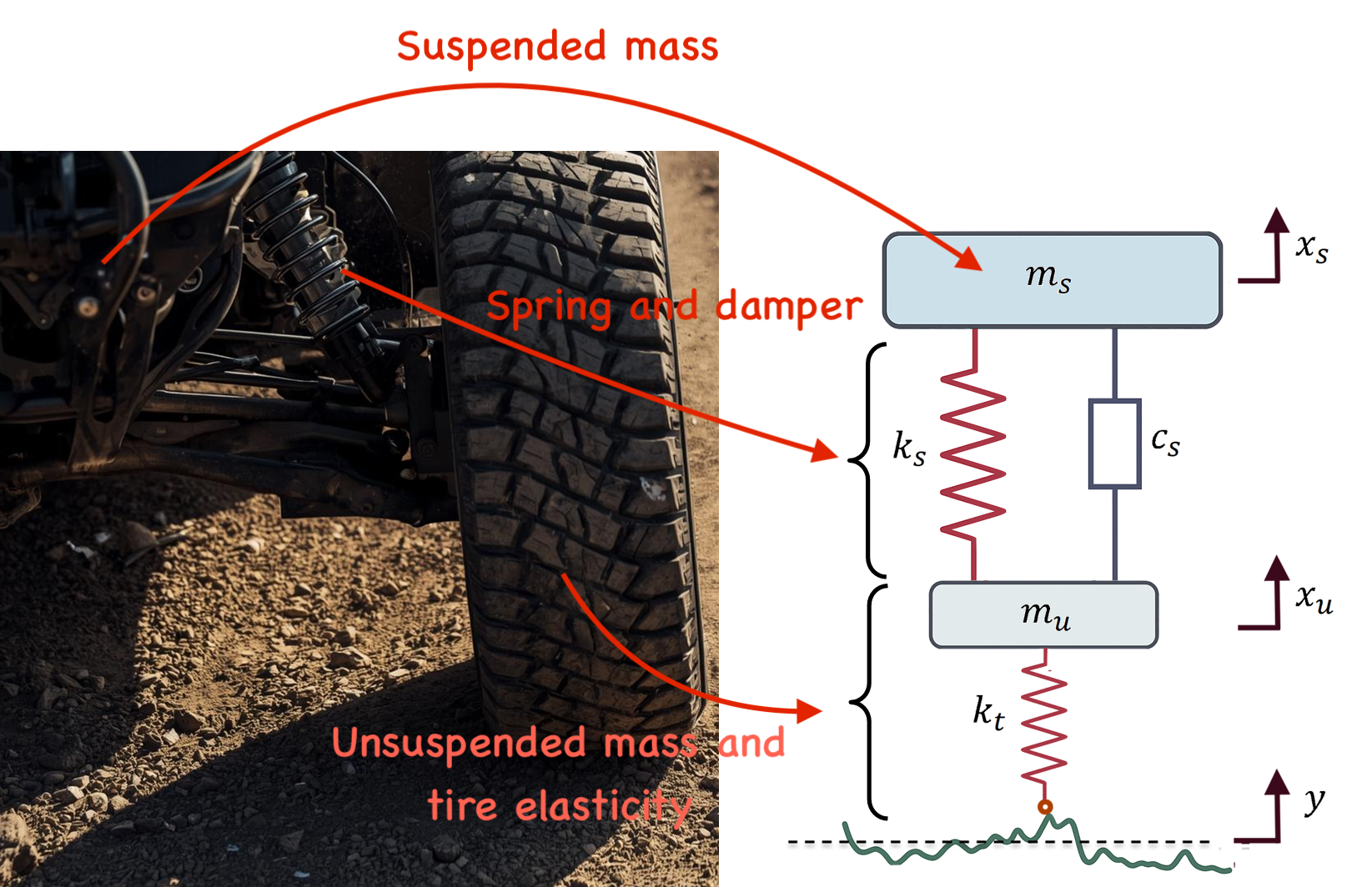}
\caption{Schematic representation of the quarter-car suspension model and its physical counterpart (AI-generated composition). The sprung mass $m_s$ represents the vehicle body, supported by the suspension spring $k_s$ and damper $c_s$. In contrast, the unsprung mass $m_{u}$ corresponds to the wheel–axle assembly interacting with the tire stiffness $k_t$. The road input $y\left(x_e(t)\right)$ excites the system, enabling the evaluation of ride comfort, road-holding performance, and transient behavior when the road profile presents sudden elevation changes. }
\label{Figure_02}
\end{figure}

The quarter-car model shown in Fig.~\ref{Figure_02} is a simplified yet widely adopted mechanical representation of a vehicle’s vertical dynamics. It captures the suspension system's essential behavior by modeling one-fourth of the vehicle’s mass, including one wheel and its associated suspension components. This reduced-order approach allows the study of suspension performance under various road excitations and facilitates control design and optimization, particularly when computational efficiency is required.

The sprung mass $m_s$ represents the portion of the vehicle supported by the suspension (i.e., the chassis), while the unsprung mass $m_u$ represents the wheel and axle assembly. The suspension system is characterized by its stiffness $k_s$ and damping coefficient $c_s$, and the tire is modeled as a spring with stiffness $k_t$. The vertical displacements of the sprung and unsprung masses are denoted by $x_s(t)$ and $x_u(t)$, respectively. At the same time, $y\left(x_e(t)\right)$ represents the road profile height, $x_e(t)$ is the tire position at time $t$, and $g$ accounts for externally applied vertical forces, such as gravitational loading or equivalent inertial terms.

The vertical dynamics of the quarter-car system are written in standard state–space form as
\begin{equation}
\begin{cases}
\dot{\bm{X}}(t) = [A(t)]\,\bm{X}(t) + [B(t)]\,\bm{U}(t) \\
\bm{Y}(t) = [C(t)]\,\bm{X}(t) + [D(t)]\,\bm{U}(t)
\end{cases}
\label{eq:ss_compact}
\end{equation}
where $\bm{X}(t) \in \mathbb{R}^4$ is the state vector; $\bm{U}(t) \in \mathbb{R}^2$ is the input vector; $\bm{Y}(t) \in \mathbb{R}^2$ is the output vector; $[A(t)] \in \mathbb{R}^{4 \times 4}$ is the system state-space matrix; $[B(t)] \in \mathbb{R}^{4 \times 2}$ is the input matrix; $[C(t)] \in \mathbb{R}^{2 \times 4}$ is output matrix; and $[D(t)] \in \mathbb{R}^{2 \times 2}$ is the feedthrough matrix. 

The state vector is defined as
\begin{equation}
\bm{X}(t) =
\begin{bmatrix}
x_u(t) & \dot{x}_u(t) & x_s(t) & \dot{x}_s(t)
\end{bmatrix}^{T},
\label{eq_state_variables}
\end{equation}
while the input vector is given by
\begin{equation}
\bm{U}(t) =
\begin{bmatrix}
y\left(x_e(t)\right) & g
\end{bmatrix}^{T},
\end{equation}
and the output vector read as
\begin{equation}
\bm{Y} =
\begin{bmatrix}
a_s(t) & f_t(t)
\end{bmatrix}^{T},
\end{equation}
where $a_s(t)=\ddot{x}_s(t)$ is the vertical acceleration of the sprung mass and $f_t(t)$ is the tire–road contact force. This formulation links suspension parameters to performance-relevant outputs, providing a transparent basis for scenario simulation and optimization in terms of the normalized design variables introduced in Sec.~\ref{subsec24}.

The system matrices in Eq.~(\ref{eq:ss_compact}) are given by
\begin{equation}
[A(t)] =
\begin{bmatrix}
0 & 1 & 0 & 0 \\
-\dfrac{k_t(t) + k_s}{m_u} & -\dfrac{c_s(t)}{m_u} & \dfrac{k_s}{m_u} & \dfrac{c_s(t)}{m_u} \\
0 & 0 & 0 & 1 \\
\dfrac{k_s}{m_s} & \dfrac{c_s(t)}{m_s} & -\dfrac{k_s}{m_s} & -\dfrac{c_s(t)}{m_s}
\end{bmatrix},
\qquad
[B(t)] =
\begin{bmatrix}
0 & 0 \\
\dfrac{k_t(t)}{m_u} & -1 \\
0 & 0 \\
0 & -1
\end{bmatrix},
\end{equation}

\begin{equation}
[C(t)] =
\begin{bmatrix}
\dfrac{k_s}{m_s} & \dfrac{c_s(t)}{m_s} & -\dfrac{k_s}{m_s} & -\dfrac{c_s(t)}{m_s} \\
k_t(t) & 0 & 0 & 0
\end{bmatrix},
\qquad
[D(t)] =
\begin{bmatrix}
0 & -1 \\
- k_t(t) & 0
\end{bmatrix}.
\end{equation}

This compact formulation highlights the structural properties of the quarter-car dynamics. It provides a convenient framework for analyzing ride comfort, road-holding capability, and transient response under different excitation scenarios. For the purpose of explaining the emergence of damping asymmetry as an optimal design feature, the quarter-car model constitutes the simplest mechanical system in which comfort, tire–road contact, and transient dynamics can be simultaneously quantified.

Effects such as dry friction, detailed damper hysteresis, nonlinear spring behavior, suspension kinematics, and full-vehicle pitch and roll dynamics are intentionally neglected. While these effects are relevant for detailed design, their inclusion would introduce model-specific parameters that hinder generalization and obscure the fundamental mechanisms governing asymmetric damping effectiveness.

\subsection{Suspension and tire characteristics} 
\label{subsec22}

The suspension system comprises the main spring, the shock absorber, and the tire’s elastic contribution. The suspension spring is modeled as a linear element with stiffness $k_s$, while the damper exhibits an asymmetric nonlinear behavior. The damping coefficient depends on the relative velocity between the sprung and unsprung masses
\begin{equation}
c_s(t) =
\begin{cases} 
c_{p}   \ \ \ \ {\dot x}_{s}(t) - {\dot x}_{u}(t) \geq 0 \\ 
c_{n}   \ \ \ \ {\dot x}_{s}(t) - {\dot x}_{u}(t)<0
\end{cases}
\label{eq_cs_asymm}
\end{equation}
where $c_p$ and $c_n$ denote the rebound and compression damping coefficients, respectively, typically satisfying $c_p > c_n$. This asymmetry reflects the larger damping forces required during rebound motion.

This asymmetric damping law constitutes the only directional asymmetry intentionally introduced in the present model. All other system components—including the suspension spring and the tire—are modeled as directionally symmetric elements. This deliberate modeling choice isolates and examines the role of damping asymmetry independently, ensuring that any observed directional effects in the system response arise solely from the damper characteristics rather than from additional geometric or constitutive asymmetries.

Tire stiffness is modeled as a strongly nonlinear element due to the possibility of wheel lift-off
\begin{equation}
k_t(t) =
\begin{cases} 
k_{t}   \ \ \ \ y(t) - x_{u}(t) \geq 0 \\ 
0       \ \ \ \ \ y(t) - x_{u}(t)<0
\end{cases}
\label{eq_kt_nonlinear}
\end{equation}
indicating that once the wheel loses contact with the road surface ($y(t)-x_u(t) < 0$), the transmitted vertical force becomes zero. This condition is important for safety analysis, as loss of contact implies momentary loss of steering, braking, and traction capabilities.

Equations (7) and (8) introduce nonlinear characteristics into the suspension model because the damping coefficient and the tire stiffness depend on the system states. As a consequence, the parameters $c_s$ and $k_t$ appearing in the state-space representation of Eq.~(1) become functions of the state variables, which implies that the system matrix is no longer constant but state-dependent during the simulation.

For numerical implementation, the piecewise definitions of these parameters are replaced by smooth representations given by
\begin{equation}
\tilde{c}_{s}(t) =
\frac{c_p + c_n}{2} +
\frac{c_p - c_n}{2}
\cdot
\frac{\dot{x}_s(t)-\dot{x}_u(t)}
{\sqrt{\left[\dot{x}_s(t)-\dot{x}_u(t)\right]^2}+\varepsilon}
\label{eq_csef}
\end{equation}
\
\begin{equation}
\tilde{k}_{t}(t) =
\frac{k_t}{2}
\left(
1+
\frac{y(t)-x_u(t)}
{\sqrt{\left[y(t)-x_u(t)\right]^2}+\varepsilon}
\right) \, ,
\label{eq_ktef}
\end{equation}
where $\varepsilon=10^{-6}$ is a regulatization parameter.

These expressions employ smooth approximations of the discontinuous functions to avoid the numerical difficulties typically associated with discontinuities when integrating the nonlinear system with standard ODE solvers.

\subsection{Road excitation} 
\label{subsec23}

A realistic evaluation of suspension dynamics requires representative excitation inputs that reproduce both the continuous irregularities of real pavements and the short-duration disturbances that produce transient vehicle responses \cite{cunha2017quantification}. For steady-state analysis, road roughness is commonly modeled as a stochastic process defined through the spatial power spectral density (PSD), as standardized by ISO 8608 \cite{ISO8608,DoddsRobson1973Roughness}. Beyond providing realism, the ISO 8608 framework ensures that the excitation statistics correspond to standardized road classes that are implicitly assumed in empirical suspension design rules and tuning practices. For transient analysis, deterministic inputs, such as step and bump profiles, are used to examine oscillatory decay and settling characteristics following abrupt perturbations.

According to ISO 8608, each road class is characterized by the PSD value $G_d(n_0)$ at a reference spatial frequency $n_0 = 0.1\ \text{m}^{-1}$, assuming a spectral decay slope $w=2$ \cite{gorges2018road,dharankar2017numerical,schiehlen2009vehicle}. The PSD for a generic spatial frequency $n$ is given by:

\begin{equation}
G_d \left( n \right)=G_d \left( n_0 \right) \left( \cfrac{n}{n_0} \right )^ {-w} \, , \ \ n_0=0.1  \ \ \text{and} \ \ \ 0.011  <n< 2.83 .
\label{eq_Gd}
\end{equation}

In this context, Table~\ref{tab_road_class} summarizes the properties of the road categories defined by ISO 8608, which provide the basis for generating the random inputs used in the simulations. These parameters characterize the spatial variability and roughness intensity of the profiles, forming the foundation for the subsequent numerical analysis. 
\begin{table}[h!]
\caption{Properties of the road classes according to ISO 8608 \cite{gorges2018road}}
\label{tab_road_class}
\begin{tabular}{>{\centering\arraybackslash}m{1.5cm} >{\centering\arraybackslash}m{1.5cm} >{\centering\arraybackslash}m{7cm}}
\toprule
ISO Class & $G_d(n_0) \ \left( 10^{-6} \text{m}^3\right)$ & Description \\
\hline
A & 16 & Airport runways and superhighways \\
B & 64 & Normal pavements \\
C & 256 & Unpaved roads and damaged pavements \\
D & 1024 & Rough unpaved roads \\
E & 4096 & Enduro tracks \\
F &  16384 & Off-road tracks \\
G &  65536 & Rough off-road tracks \\
H &  262144 & Simulation purpose only \\
\bottomrule
\end{tabular}
\end{table}

A pseudorandom road profile is synthesized by summing $N$ sinusoidal components, with amplitudes derived from the PSD and whose random phases are uniformly distributed \cite{gorges2018road,lenkutis2021road, shen2021optimal}, i.e.,
\begin{equation}
y\left(x_e(t)\right)= \sum_{i=1}^{N} A_i \sin \left( {2 \, \pi \, n_i \, x_e(t) + \Phi_i} \right) \ \ \ \Phi_i \sim \mathcal{U}(0, 2 \pi) \, ,
\label{eq_rough_road}
\end{equation}
where the amplitudes are given by
\begin{equation}
A_i=\sqrt{G_d \left( n_i \right) \Delta n} \, .
\label{eq_Ai}
\end{equation}

It is worth noting that, although the road profile is generated to match a prescribed PSD, the random phases $\Phi_i$ introduce variability in each realization of the surface. As a consequence, different pseudorandom road profiles, while statistically equivalent, may lead to slightly different dynamic responses of the vehicle. This variability affects the evaluation of performance metrics such as the sprung mass acceleration and the tire--ground contact force, and may result in small variations in the optimal suspension parameters. Therefore, the optimization results should be interpreted in a statistical sense, as they depend on a specific realization of the stochastic road profile. Nevertheless, this variability does not compromise the overall conclusions of the study, since the observed trends and relative performance of the suspension configurations remain consistent across different realizations. For increased robustness, it is recommended to perform the analysis over multiple realizations and to evaluate the convergence of the optimal parameters and performance indices in terms of their mean and dispersion.

\begin{figure}
\centering
\includegraphics[width=130mm]{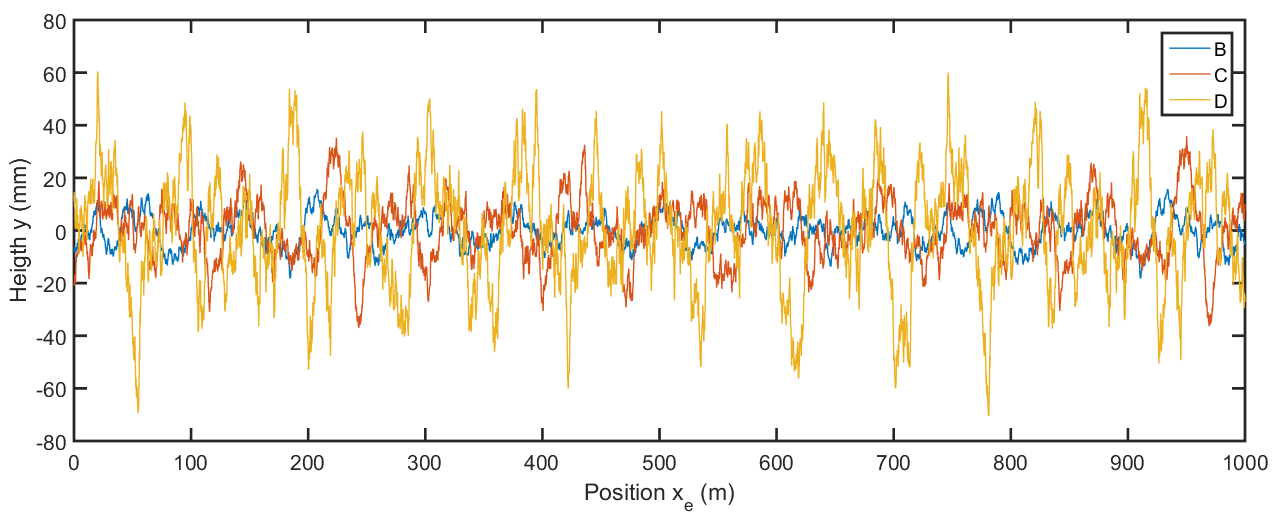}
\caption{Simulated vertical road profiles for ISO 8608 classes B, C, and D, generated from the corresponding spatial power spectral density functions. The curves illustrate progressively higher surface roughness as the class index increases and are used to evaluate the suspension’s permanent (steady-state) response under continuous excitation.}
\label{Figure_03}
\end{figure}

Figure~\ref{Figure_03} shows the synthesized road profiles for ISO classes B, C, and D, illustrating the increase in surface roughness and excitation intensity as the class index rises. These stochastic profiles are used to evaluate the suspension’s permanent, steady-state response under continuous broadband excitation.

\begin{figure}
\centering
\includegraphics[width=130mm]{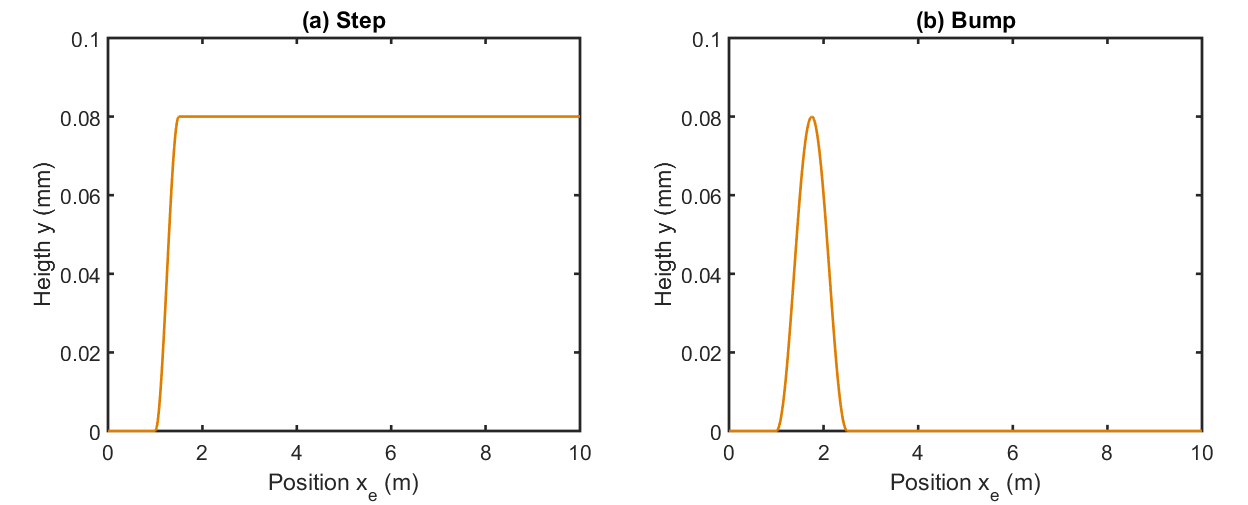}
\caption{Road excitation functions used to assess the suspension’s transient response: (a) a step input, representing an abrupt change in road elevation, and (b) a bump input, modeling a localized disturbance. Both profiles are applied to analyze the vehicle’s short-term dynamic response, including oscillation decay and settling time after sudden perturbations.}
\label{Figure_04}
\end{figure}

In addition to stochastic roughness, the deterministic excitation functions shown in Fig.~\ref{Figure_04} are used to evaluate the suspension’s transient behavior. A step input represents an abrupt and sustained change in road elevation, allowing the analysis of immediate displacement response, overshoot, and settling time. A bump input models a localized geometric disturbance, typically used to assess oscillation decay, peak accelerations, and the interaction between the damping characteristics and natural frequency during short-duration events. These profiles complement the stochastic model by isolating specific dynamic phenomena associated with sudden changes in road elevation. 

\subsection{Simulation setup and operating conditions} 
\label{subsec24}

To support the analysis and optimization of suspension configurations, all system parameters are expressed as normalized quantities that characterize the quarter-car model's dynamic behavior. The suspension is therefore parameterized using the natural frequency $f_n$ and the damping ratio $\zeta_i$, which provide a dimensionless representation suitable for comparing different vehicle classes. 

These quantities are defined as
\begin{equation}
f_n=\cfrac{1}{2 \pi} \sqrt{\cfrac{k_s}{m_s}}\\[6pt]
\label{suspension_parameters1}
\end{equation}
and
\begin{equation}
\zeta_i=\cfrac{c_i}{2 \sqrt{k_s \, m_s}} \qquad i=p \ \text{or} \ n,
\label{suspension_parameters2}
\end{equation}
where $c_p$ denotes the rebound damping coefficient and $c_n$ the compression damping coefficient. This formulation enables scalable characterization of the suspension, facilitating investigation of damping asymmetry and stiffness variations across different operating conditions.

This normalization enables the identification of scenario-dependent trends independent of specific vehicle realizations, thereby supporting the interpretation of asymmetric damping ratios as structural rather than vehicle-specific design features.

Before performing the optimization, numerical simulations are conducted to compute the resulting accelerations and tire forces for the selected road profiles. To ensure consistent comparison across different suspension configurations, all simulations are performed under a fixed set of operating conditions that define the reference parameters of the quarter-car model—such as masses, tire stiffness, road roughness classification, and vehicle speed. Table~\ref{tab_parameters} summarizes the parameters adopted in the simulation and optimization stages.

\begin{table}[h]
\begin{center}
\begin{minipage}{770pt}
\caption{Key parameters adopted in the suspension simulation and optimization process, including vehicle masses, tire stiffness, road roughness classification, and operating conditions.}
\label{tab_parameters}
\begin{tabular}{@{}cccc@{}}
\toprule
Parameter & Units & Value & Description \\
\midrule
$m_s$ & kg & 250 (light) / 500 (mid-heavy) & Sprung mass \\
$m_u$ & kg & 40 (light) / 50 (mid-heavy) & Unsprung mass \\
$k_t$ & N/m & 200000 (light) / 250000 (mid-heavy) & Tire stiffness\\
ISO Class & -- & B, C and D & Road roughness\\
$L$ & m & 1000 & Road length\\
$v$ & m/s & 20 and 40 & Travel speed\\
\bottomrule
\end{tabular}
\end{minipage}
\end{center}
\end{table}

The adopted parameters do not correspond to a specific commercial vehicle but represent typical values for light and medium–heavy vehicles reported in the literature. These parameters define a reference operating condition for the simulations; however, they appear explicitly in the dynamic model and may be varied to investigate other operating scenarios. In this way, the proposed formulation can be applied to different vehicle configurations and operating conditions in order to determine combinations of vehicle mass and travelling speed that maintain $R_{f_t}$ and $\sigma_{a_w}$ within acceptable limits for a given road roughness class.

\section{Analysis of optimal suspension configurations} 
\label{sec3}

This section analyzes suspension performance through the lens of optimization-based design. Rather than prescribing damping asymmetry through empirical rules, the objective here is to quantify how ride comfort, road holding, and transient response depend on suspension parameters, and to determine which configurations emerge as optimal under representative operating scenarios. By combining standardized performance metrics with a formal optimization framework, this section provides the analytical bridge between commonly adopted empirical tuning practices and their underlying dynamical justification.

\subsection{Ride comfort assessment} 
\label{subsubsec31}

Ride comfort is a central objective in suspension design, as it directly affects driver fatigue, passenger well-being, and overall vehicle safety. Human sensitivity to whole-body vibration varies with direction, amplitude, frequency, and exposure duration \cite{els2005applicability, griffin2007discomfort}. To standardize the evaluation of these effects, ISO 2631-1:1997 \cite{ISO2631_1_1997,Griffin1990Handbook} establishes procedures for quantifying vibration severity and classifying the associated discomfort levels.

The human body is not equally sensitive to all frequencies. Vertical vibrations between 4 and 8 Hz coincide with the primary resonance of the abdomen and thoracic cavity, resulting in increased discomfort and degraded cognitive performance. To model this frequency-dependent sensitivity, ISO 2631 defines weighting filters that enhance or attenuate specific frequency bands \cite{oh2017digital}. For vertical seated occupants, the $W_k$ filter is appropriate, emphasizing frequencies where the human body is most sensitive.

The weighted acceleration signal $a_w(t)$ is used to compute the corresponding RMS value,

\begin{equation}
\sigma_{a_w}=
\sqrt{\frac{1}{T} \int_0^T a_w(t)^2 \, dt},
\label{eq_acceleration}
\end{equation}
where $T$ is the exposure duration. The scalar $\sigma_{a_w}$ represents the severity of vibration and forms the basis for comfort classification. 

Table~\ref{tab_comfort} summarizes the comfort ranges defined in ISO 2631-1:1997. These overlapping intervals capture the intrinsically subjective nature of human perception.

\begin{table}[h]
\caption{Expected comfort reactions to vibration environments according to ISO 2631. Ranges overlap because descriptors reflect subjective perception.}
\label{tab_comfort}
\begin{tabular}{>{\centering\arraybackslash}m{5cm} >{\centering\arraybackslash}m{5cm}}
\toprule
RMS weighted acceleration ($\mathrm{m/s^2}$) & Comfort level \\
\midrule
$<$ 0.315 & not uncomfortable \\
0.315–0.63 & a little uncomfortable \\
0.5–1.0 & fairly uncomfortable \\
0.8–1.6 & uncomfortable \\
1.25–2.5 & very uncomfortable \\
$>$ 2.0 & extremely uncomfortable \\
\bottomrule
\end{tabular}
\end{table}

Ride-comfort evaluation based on ISO 2631 thus provides an objective connection between measured vehicle dynamics and perceived human response. It enables a rigorous and standardized comparison of suspension configurations, forming a quantitative basis for comfort-oriented optimization and trade-off analysis.

\subsection{Tire–ground contact force} 
\label{subsubsec32}

The tire–ground contact force is essential for maintaining vehicle control, as it governs the achievable traction, steering capability, and braking performance. Under dynamic excitation, this force fluctuates due to road irregularities, suspension motion, and the inertia of the sprung and unsprung masses. Excessive fluctuations may lead to partial or total loss of contact, compromising stability and safety.

A common statistical measure of contact-force variability is the standard deviation of the tire force, $\sigma_{f_t}$ \cite{uys2006criteria, uys2007suspension, schiehlen2009vehicle}. To express this quantity in a normalized and dimensionless form, the contact-force ratio is defined as
\begin{equation}
R_{f_t}=\frac{\sigma_{f_t}}{(m_s + m_u)\, g},
\label{eq_force_variation}
\end{equation}
where $g$ is the gravitational acceleration. The parameter $R_{f_t}$ quantifies the relative fluctuation of the tire--ground contact force with respect to the static load. This metric is particularly suitable for rough road conditions, where the variations of the contact force tend to follow an approximately Gaussian distribution. Under this assumption, values of $R_{f_t}$ lower than $0.3$ indicate that the fluctuations of $f_t$ remain sufficiently bounded such that zero contact force is unlikely to occur, meaning that tire--ground contact is maintained throughout the operation. In the present study, values of $R_{f_t}$ up to $0.25$ are considered acceptable in order to ensure a safe margin against loss of tire--ground contact.

In addition, $R_{f_t}$ is also widely adopted as an indicator of road-friendliness, since larger dynamic tire force fluctuations are directly associated with increased pavement damage. Therefore, limiting $R_{f_t}$ not only contributes to maintaining continuous tire--ground contact but also helps mitigate the dynamic loading transmitted to the pavement, reducing infrastructure deterioration, as discussed in \cite{sun2002optimum,misaghi2021impact}.

\begin{figure}
\centering
\includegraphics[width=130mm]{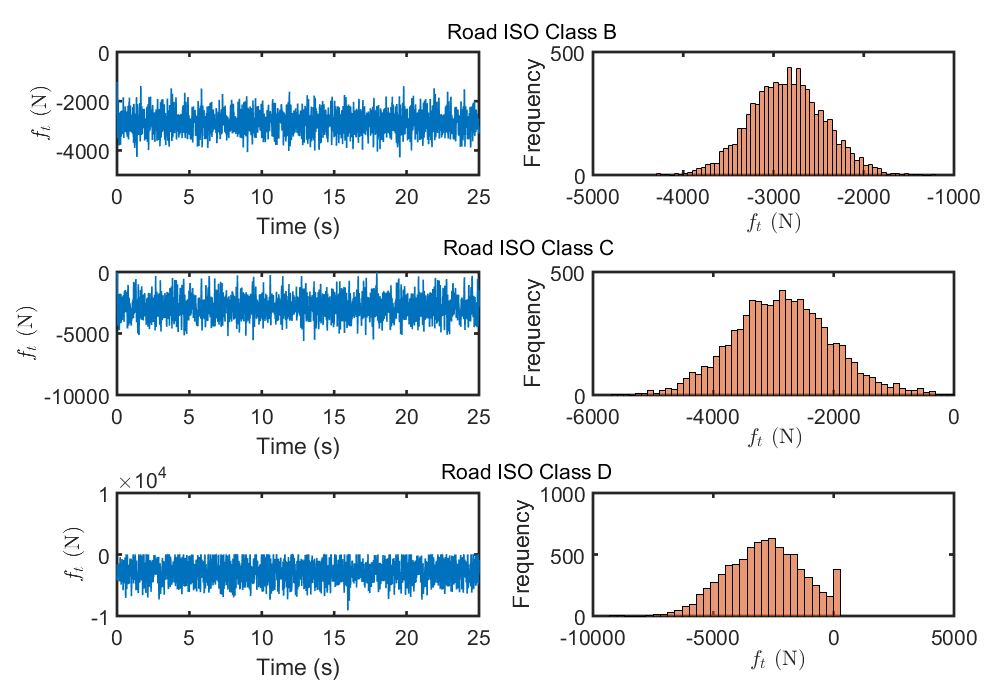}
\caption{Resulting tire–ground contact force $f_t$ for a light vehicle with $f_n = 1.5$ Hz and $c_p = c_n = 0.3$, traveling at $v = 40$ m/s on roads of different classes.}
\label{Figure_Rft_definition}
\end{figure}

Figure~\ref{Figure_Rft_definition} illustrates the tire--ground contact force $f_t$ for a light vehicle with $f_n = 1.5$ Hz and $c_n = c_p = 0.3$, traveling at a velocity $v = 40$ m/s on roads of different ISO roughness classes. It can be observed that for an ISO Class B road the value obtained is $R_{f_t} = 0.15$, for which no loss of tire--ground contact occurs. For an ISO Class C road the value increases to $R_{f_t} = 0.30$, representing a limiting condition where the contact force approaches zero. For an ISO Class D road the value reaches $R_{f_t} = 0.53$, indicating frequent occurrences of $f_t = 0$, which correspond to repeated losses of tire--ground contact. These results illustrate how increasing road roughness leads to larger values of $R_{f_t}$ and consequently to a higher probability of loss of tire--ground contact.

This metric complements passenger-comfort indicators, such as $\sigma_{a_w}$. While comfort indices assess the human experience, $R_{f_t}$ evaluates mechanical adherence to the ground, making it fundamental for identifying suspension configurations that balance comfort and vehicle stability.

\subsection{Transient response} 
\label{subsubsec33}

In vehicle dynamics analysis, transient behavior describes how the system dissipates disturbances and returns to steady operating conditions. For vertical dynamics, such disturbances are typically induced by abrupt variations in road elevation, including localized bumps or step-like irregularities, which can be represented as transient inputs acting on the suspension system. The response following these excitations provides valuable insight into suspension performance, particularly through the evaluation of the settling time, defined as the duration required for the oscillatory motion of the sprung mass to decay to an acceptably small amplitude.

A shorter settling time reflects the suspension’s ability to suppress oscillations efficiently, allowing the vehicle body to stabilize rapidly after a disturbance. This characteristic improves ride comfort and contributes to road-holding by enabling the tire–ground forces to reestablish stable levels in a shorter time. Conversely, prolonged settling times are associated with sustained body motion and may impair stability, especially at higher speeds. It is worth noting that transient response analysis is also widely employed in other domains of vehicle dynamics; for example, the lateral response to a step steering input is commonly used as an indicator of handling performance \cite{uys2006criteria}.

\begin{figure}
\centering
\includegraphics[width=130mm]{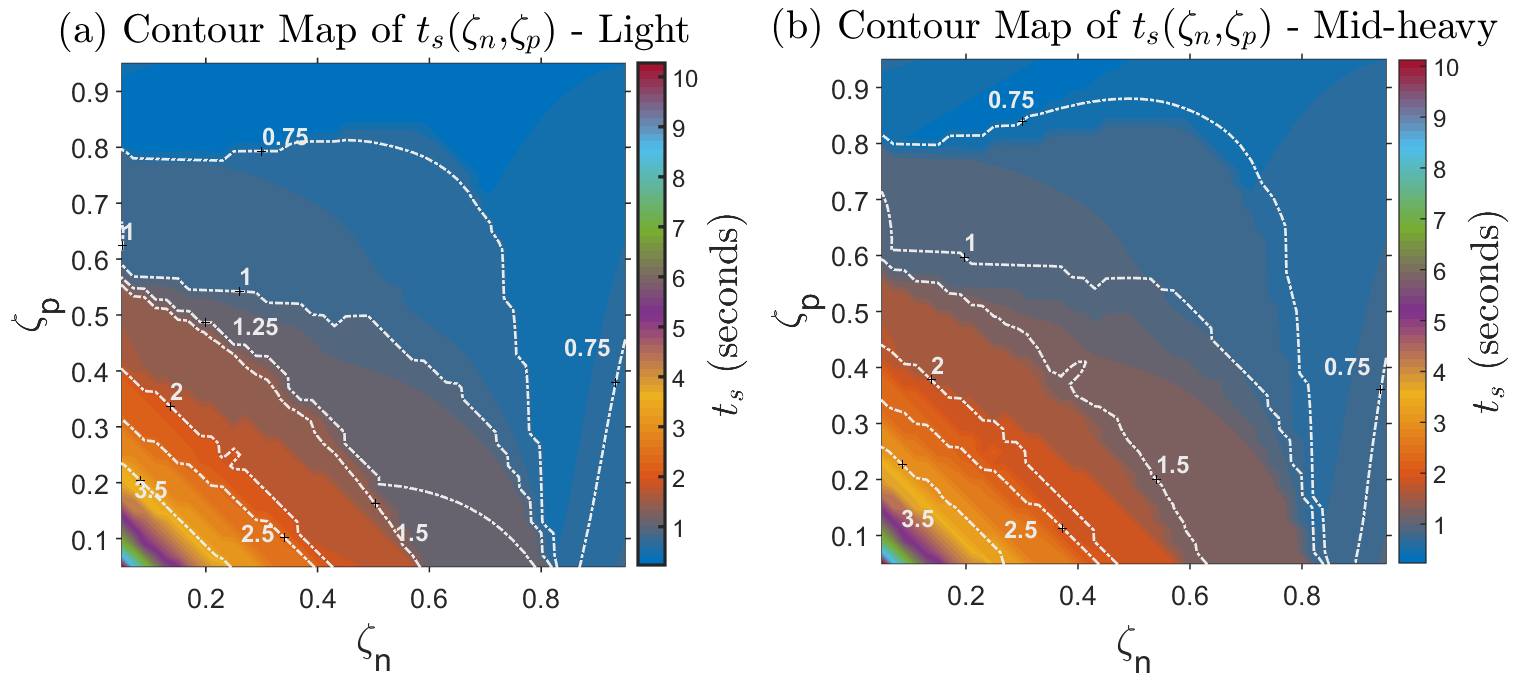}
\caption{Contour maps of settling time for (a) light vehicles and (b) mid-heavy vehicles as functions of the compression and rebound damping coefficients. Vehicle parameters are defined in Tab.~\ref{tab_parameters},  assuming a suspension natural frequency of $f_n=1.5 \text{ Hz}$. }
\label{Figure_Contour_Ts}
\end{figure}

From a design perspective, damping parameters are tuned to achieve a transient response that balances comfort and handling: the decay should be fast enough to ensure stability and compliance, yet not so rapid that it introduces harshness. In passenger vehicles, primary ride modes typically exhibit settling times on the order of 1–1.5 seconds, whereas lateral stability controllers often target even faster decay for yaw dynamics.

A practical advantage of analyzing the sprung-mass displacement following a bump excitation lies in its experimental accessibility. In practice, this quantity can be obtained from acceleration measurements using accelerometers through double integration procedures. Although such experimental reconstruction is not performed in the present study, this observation highlights the relevance of displacement-based analysis for potential validation and experimental applications.

\subsection{Formulation of the optimal design problem} 
\label{subsec34}

Based on the performance metrics introduced in the previous subsections—namely ride comfort quantified by the weighted RMS acceleration and road-holding assessed through the tire--ground contact force ratio—the suspension design problem can be formulated as a multi-objective optimization task \cite{peng2021multiobjective,he2019commercial,goga2012optimization,shirahatti2008optimal}. Ride comfort and safety tend to deteriorate under critical operating conditions that combine higher vehicle speeds and/or roads with higher ISO roughness classes, as these conditions lead to larger accelerations of the sprung mass and higher values of $R_{f_t}$.

Regarding ride comfort, Table~\ref{tab_comfort} presents reference values that can be used as limits for the occupants of the vehicle. In such situations, the acceleration levels may be reduced by adopting a softer suspension. Concerning safety, the acceptable limit was defined as $R_{f_t} \leq 0.25$. These values also tend to increase with higher vehicle speeds and rougher roads, but they can likewise be reduced by using a softer suspension.

The inherent conflict between these criteria \cite{els2007ride,Torun2025RailComfortDerailmentTradeoffJVETech}, particularly under severe road excitation, requires a systematic approach that can explore trade-offs and identify feasible compromises among comfort, safety, and transient performance. Thus, the suspension design problem is addressed under three optimization scenarios: (i) pure comfort optimization, (ii) pure road-holding optimization, and (iii) constrained trade-off optimization, in which one performance metric is minimized while the other is driven toward a prescribed target value.

To formally state the design problem, first let the vector of design variables be defined as
\begin{equation}
\boldsymbol{\theta} =
\begin{bmatrix}
f_n & \zeta_p & \zeta_n
\end{bmatrix}^{T},
\end{equation}
which parameterizes the suspension stiffness and asymmetric damping characteristics. Then, define the admissible design space as the feasible set
\begin{equation}
\Theta = \left\{ \boldsymbol{\theta} \in \mathbb{R}^3 \; \middle| \;
1.0 \leq f_n \leq 2.0,\;
0 \leq \zeta_p \leq 1,\;
0 \leq \zeta_n \leq 1
\right\}.
\end{equation}

Within the bounds defined by $\Theta$, the goal is to find designs that optimally balance ride comfort and tire–ground adherence. This formulation makes explicit how asymmetric damping ratios arise as optimal design variables rather than imposed heuristics.

Therefore, the suspension design problem is formulated as the following parameter optimization problem
\begin{equation}
\boldsymbol{\theta}^\ast
=
\arg \min_{\boldsymbol{\theta} \in \Theta}
J(\boldsymbol{\theta}),
\label{eq:opt_problem}
\end{equation}
with the scalar objective function defined as
\begin{equation}
J(\boldsymbol{\theta}) =
A_s \, \left| \sigma_{a_w}(\boldsymbol{\theta}) - \sigma_{a_w}^{\mathrm{ref}} \right|
+
A_f \, \left| R_{f_t}(\boldsymbol{\theta}) - R_{f_t}^{\mathrm{ref}} \right|.
\label{eq_objective_function}
\end{equation}

The objective function defined in Eq.~(\ref{eq_objective_function}) is evaluated for prescribed operating conditions, namely the vehicle natural frequency $f_n$, the vehicle velocity $v$, and the road roughness profile $y(x_e)$. Under these conditions, the suspension parameters are determined by minimizing the deviation between the performance metrics and their respective reference values.

When $A_s=1$ and $A_f=0$, the optimization reduces to a single-objective problem in which the suspension configuration that minimizes the weighted RMS acceleration is obtained, yielding the minimum value $\sigma_{a_w}^{\min}$. For this configuration, the corresponding tire–ground contact force ratio is $R_{f_t}(\sigma_{a_w}^{\min})$.

Conversely, when $A_f=1$ and $A_s=0$, the optimization focuses exclusively on maximizing road-holding, resulting in the suspension configuration that minimizes the contact-force ratio $R_{f_t}^{\min}$. In this case, the corresponding acceleration level is $\sigma_{a_w}(R_{f_t}^{\min})$. The suspension obtained in the first case is typically softer than the one obtained in the second case, reflecting the well-known trade-off between ride comfort and road-holding.

Intermediate suspension configurations can be obtained by prescribing reference values for one performance metric while minimizing the deviation of the other. For instance, by specifying a target value $R_{f_t}^{\mathrm{ref}}$ between $R_{f_t}(\sigma_{a_w}^{\min})$ and $R_{f_t}^{\min}$ and selecting $A_s=1$, $A_f=100$, and $\sigma_{a_w}^{\mathrm{ref}}=0$, the optimization yields the suspension configuration that minimizes acceleration while satisfying the desired road-holding level. Alternatively, by prescribing a reference acceleration $\sigma_{a_w}^{\mathrm{ref}}$ between the values obtained in the two extreme configurations and selecting $A_s=100$, $A_f=1$, and $R_{f_t}^{\mathrm{ref}}=0$, the optimization determines the suspension that minimizes $R_{f_t}$ while achieving the desired comfort level.

The weighting factors adopted to obtain these intermediate solutions are reported in Table~\ref{tab_objective_function}. Their numerical values were selected empirically in order to enforce the desired dominance of one performance metric over the other while preserving the numerical stability of the optimization process.

Finally, the optimized solutions must be evaluated against the admissible limits discussed previously. If the resulting values exceed acceptable thresholds for a given road roughness class, the feasible operating conditions may require reducing the vehicle speed and/or considering a heavier vehicle configuration.

\begin{table}[h]
\caption{Objective function parameters for predifened values for $m_s, m_u, k_f \ \text{and} \ f_n$.}
\label{tab_objective_function}
\begin{tabular}{>{\centering\arraybackslash}m{4cm} >{\centering\arraybackslash}m{1cm} >{\centering\arraybackslash}m{1cm} >{\centering\arraybackslash}m{2cm} >{\centering\arraybackslash}m{2cm}}
\toprule
Optimization objective & $A_s$ & $A_f$ & $\sigma_{a_w}^{\text{ref}}$ & $R_{f_t}^{\text{ref}}$ \\
\midrule
Min $\sigma_{a_w}$ & 1 & 0 & 0 & 0\\
Min $R_{f_t}$ & 0 & 1 & 0 & 0\\
Min $\sigma_{a_w}$ given $R_{f_t}$ & 1 & 100 & 0 & desired value \\
Min $R_{f_t}$ given $\sigma_{a_w}$ & 100 & 1 & desired value & 0\\
\bottomrule
\end{tabular}
\end{table}

This optimization problem is intrinsically non-convex. This non-convexity does not arise solely from the choice of objective function, but primarily from the nonlinear and non-smooth dependence of the performance metrics on the design parameters through the underlying dynamical system. The mapping from suspension parameters to ride comfort, road-holding, and transient metrics involves time-domain simulations of a piecewise-defined nonlinear system, followed by nonlinear statistical post-processing, which precludes convexity and the use of gradient-based optimization methods. Schematically, we have
\[
(\zeta_p,\zeta_n,f_n)
\;\longrightarrow\;
\substack{\text{nonlinear ODE}\\ \text{simulation}}
\;\longrightarrow\;
\substack{\text{time}\\ \text{signals}}
\;\longrightarrow\;
\substack{\text{nonlinear functionals}\\ \text{(RMS, STD)}} 
\;\longrightarrow\;
\substack{\text{objective}\\ \text{function}} 
\,\, .
\]

In this context, an optimization algorithm that can handle nonlinear dynamics, asymmetric damping characteristics, and multiple, often competing, objectives is required. In the next subsection, we present an efficient stochastic metaheuristic for global optimization that is particularly well-suited to this problem.

\subsection{Non-convex optimization via the Cross-Entropy method}
\label{subsec35}

To address the challenges underlying the suspension design problem formulated above, the Cross-Entropy (CE) method \cite{rubinstein2004cross,CunhaJr2024CEopt} is adopted as the optimization algorithm. The CE method is a stochastic, population-based Monte Carlo technique \cite{kroese2011,jra2014uncertainty}, used to solve non-complex optimization problems characterized by nonlinearity, non-smoothness, and multiple local optima. These features make it particularly suitable for the present application, in which the optimization problem is inherently non-convex and the performance metrics are obtained via nonlinear time-domain simulations of a piecewise-defined dynamical system.

Conceptually, the CE method operates by iteratively sampling candidate solutions from a probability distribution defined over the admissible design space. At each iteration, the performance of the sampled candidates is evaluated using the objective function, and a subset of high-performing solutions—often referred to as elite samples—is selected. The probability distribution is then updated to increase the likelihood of generating similar or improved solutions in subsequent iterations. Through this iterative refinement process, the distribution progressively concentrates around regions of the design space associated with near-optimal performance. Convergence is achieved when the distribution stabilizes or when improvements in the objective function become negligible.

The suitability of the CE method for problems of this nature has been demonstrated in a range of applications involving nonlinear and nonstandard dynamical systems with simulation-based objective functions.  In \cite{Cunha2021EnergyHarvestingCE}, the CE method was successfully applied to the optimization of a bistable energy harvesting device governed by nonlinear differential equations, where performance metrics were evaluated through time-domain simulations and the resulting objective landscape was strongly non-convex. Related applications include the use of CE for uncertainty quantification in mechanistic epidemic models \cite{CunhaBartonRittoEpidemicsCE} and for structural optimization problems involving dynamic response under seismic loading \cite{RodriguesCunhaBeck2023RCFrame}. More recently, the CE method has also been employed in the context of advanced fractional-order dynamical systems, including the calibration of variable-order fractional models for polymer creep \cite{TellesRibeiroCunha2025Creep,ribeiro2025advanced} and the optimal tuning of fractional-order controllers in nonlinear control problems \cite{BasilioTellesRibeiroCunha2022FracLQR}. These studies collectively illustrate the robustness and versatility of the CE method in handling complex optimization problems defined implicitly through differential equation models—integer- and fractional-order alike—reinforcing its suitability for the suspension design problem considered here.

One of the main advantages of the CE method is its algorithmic simplicity combined with a solid mathematical foundation. The method does not rely on gradient information and is therefore robust to non-smooth objective functions, discontinuities, and noisy evaluations—features that naturally arise in suspension optimization problems involving asymmetric damping and tire–ground contact conditions. In contrast to many evolutionary algorithms, such as genetic algorithms, the CE method requires a small number of intuitive control parameters, typically related to population size, elite fraction, smoothing of distribution updates, and convergence tolerance, which facilitates implementation and tuning.

In the present study, the CE method is implemented using the MATLAB package \textit{CEopt: Cross-Entropy Optimizer} \cite{CunhaJr2024CEopt} using a population size of $N_{\mathrm{CE}} = 75$ samples per iteration, with an elite proportion of $\rho = 0.1$. A smoothing parameter $\alpha = 0.8$ was adopted to ensure stable convergence, and the optimization process was limited to a maximum of 25 iterations. The search space is defined by two design variables, corresponding to the asymmetric damping ratios. This implementation provides a flexible and efficient framework for constrained, non-convex optimization and is publicly available at \url{https://ceopt.org}. Further mathematical details and algorithmic variants of the CE method can be found in \cite{rubinstein2004cross,kroese2011}.

This framework provides a systematic approach to exploring asymmetric damping and stiffness configurations and to identifying suspension designs tailored to different road conditions, while ensuring both passenger comfort and vehicle stability.

\section{Results and discussion}
\label{sec4}

This section presents and interprets the optimal suspension configurations obtained through the scenario-driven optimization framework introduced in the previous section. Rather than merely reporting numerical optima, the results are analyzed with the explicit goal of explaining when and why asymmetric damping emerges as an optimal design feature. By progressively increasing road severity and comparing vehicle classes, the results reveal how empirical damping asymmetry rules can be understood as rational responses to the coupled demands of ride comfort, road holding, and transient stability.

When a vehicle is subjected to combinations of high travel speed and severe road roughness, the suspension system is pushed to the limits of its dynamic performance. Under such operating conditions—hereafter referred to as \textit{critical situations}—the central challenge is to preserve an acceptable compromise between ride comfort and road-holding, despite the increasingly aggressive excitation imposed by the terrain. As surface irregularities intensify, the tire–ground interaction becomes more erratic, leading to amplified fluctuations in the contact force and a progressive increase in the tire–force ratio $R_{f_t}$. Once $R_{f_t}$ approaches values associated with dynamic unloading, the risk of partial or complete loss of contact rises significantly, with direct implications for vehicle safety and controllability. For this reason, maintaining $R_{f_t}<0.3$ is adopted as a practical upper bound to ensure continuous tire–ground contact.

To explore this performance boundary, a set of three representative operating scenarios with increasing severity is examined, corresponding to road classes B, C, and D, combined with different vehicle configurations and operating velocities. The results provide insight into how the feasibility of safe, comfortable operation depends on the coupled effects of vehicle mass and damping characteristics, particularly the role of damping asymmetry in mitigating tire force fluctuations under harsh excitation. These findings establish the conditions under which passive suspension systems remain viable and highlight the transition toward configurations that demand more robust design choices as operating limits are approached.

It is worth noting that the proposed formulation is not restricted to these particular values. Both vehicle mass and traveling speed appear explicitly in the dynamic model and in the excitation formulation. In particular, the vehicle speed determines the temporal characteristics of the road excitation, since the spatial road profile $y(x_e)$ is converted into a time-domain input through the relation $x_e = vt$. Consequently, different speeds lead to different excitation frequencies acting on the suspension system. Likewise, the vehicle mass directly affects the system's dynamic response through the equations of motion.

Therefore, the adopted values should be interpreted as a reference operating scenario rather than as a restriction of the method. The same optimization framework can be applied to other vehicle masses, speeds, and road roughness levels without any modification to the formulation.

In practice, variations in vehicle mass and travelling speed can be used to determine operating conditions that ensure that the performance indicators considered in this study—namely the tire–ground contact force ratio $R_{f_t}$ and the weighted RMS acceleration $\sigma_{a_w}$—remain within acceptable safety and comfort limits for a given road class. For instance, very rough roads often involve vehicles with larger effective masses and reduced speeds, as observed in off-road competitions or in military vehicles operating on desert terrain. Under such conditions, the methodology presented in this work can be used to evaluate how to adjust suspension parameters and operating conditions to maintain safe levels of road-holding and ride comfort.

\subsection{Light vehicles} \label{subsec41}

\begin{figure}[h]
\centering
\includegraphics[width=130mm]{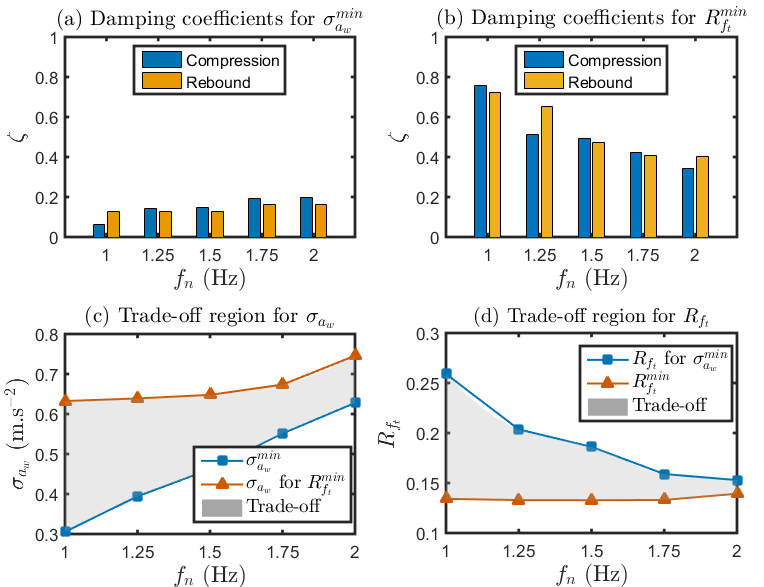}
\caption{Optimal damping coefficients for a light vehicle on a Class B road at 40 m/s, considering suspension natural frequencies of 1.0, 1.25, 1.5, 1.75, and 2.0 Hz: (a) acceleration-based optimization results; (b) tire–road contact force–based optimization results; (c) resulting acceleration levels; and (d) tire–road contact forces and corresponding trade-off regions.}
\label{Figure_Urban_B_40}
\end{figure}

The first case investigated corresponds to a light vehicle operating on a paved road classified as ISO Class B. The results obtained for this scenario are presented in Fig.~\ref{Figure_Urban_B_40}. Figure~\ref{Figure_Urban_B_40}(a) shows the damping coefficients required to minimize the weighted acceleration, whereas Fig.~\ref{Figure_Urban_B_40}(b) reports the corresponding coefficients associated with the minimization of the tire–force ratio $R_{f_t}$. As expected, the damping levels that minimize acceleration are generally lower than those required to minimize $R_{f_t}$.

An interesting observation is that the minimum values of both $\sigma_{a_w}$ and $R_{f_t}$ are, in some cases, achieved with slightly asymmetric suspension configurations, occurring in both directions. However, these deviations from symmetry are small and can be attributed to the stochastic variability of the road profile. Overall, the optimal solutions remain close to symmetric configurations, indicating no consistent or systematic requirement for damping asymmetry. This reinforces the conclusion that symmetric damping provides robust and practically suitable performance under the considered operating conditions.

\begin{figure}[h]
\centering
\includegraphics[width=130mm]{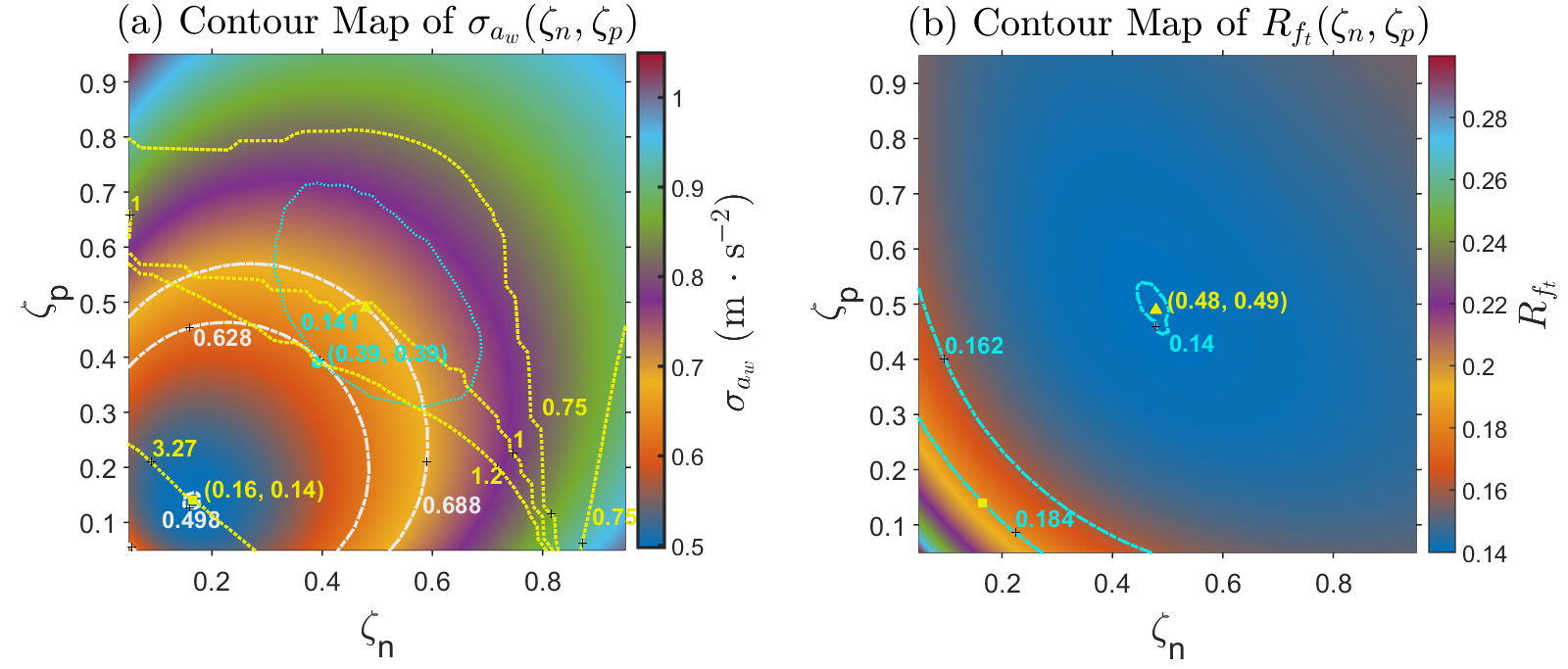}
\caption{Contour maps of suspension performance metrics for a light vehicle operating on a Class B road at a constant velocity of 40 m/s, with a suspension natural frequency of $f_n=1.5\text{ Hz}$: (a) standard deviation of the sprung-mass vertical acceleration, $\sigma_{a_w}$, used as an indicator of ride comfort; and (b) tire–road contact force variation index, $R_{f_t}$, used to assess contact stability. The contours are presented as functions of the compression and rebound damping coefficients, highlighting the influence of damping asymmetry on comfort and contact performance.}
\label{Figure_Urban_B_40_Contour_1p5}
\end{figure}

Additional insights are provided by Figs.~\ref{Figure_Urban_B_40}(c) and \ref{Figure_Urban_B_40}(d), which map the range of achievable values of weighted RMS acceleration $\sigma_{a_w}$ and tire–force ratio $R_{f_t}$ over different combinations of damping coefficients. These results allow an initial assessment of the minimum attainable acceleration for a prescribed value of $R_{f_t}$, assuming a fixed natural frequency $f_n$. To further illustrate this trade-off, Fig.~\ref{Figure_Urban_B_40_Contour_1p5} presents contour plots of $\sigma_{a_w}$ in Fig.~\ref{Figure_Urban_B_40_Contour_1p5}(a) and of $R_{f_t}$ in Fig.~\ref{Figure_Urban_B_40_Contour_1p5}(b) for $f_n=1.5 \text{ Hz}$.

Figure~\ref{Figure_Urban_B_40_Contour_1p5}(a) shows that optimizing solely for ride comfort leads to a settling time of $t_s=3.27 \text{ s}$, which can be considered excessively long and therefore undesirable from a dynamic stability perspective. In contrast, optimization focused on minimizing $R_{f_t}$ yields a significantly shorter settling time of $t_s=1.0 \text{ s}$, with corresponding values of $\sigma_{a_w}=0.688 \text{ m/s$^2$}$ and $R_{f_t}=0.140$. Although further reductions in acceleration can be achieved at the expense of increased $R_{f_t}$, the same contour plot indicates that constraining the solution to $R_{f_t}=0.141$ allows the acceleration to be reduced to $\sigma_{a_w}=0.628 \text{ m/s$^2$}$, with a moderate increase in settling time to $t_s=1.20 \text{ s}$, which remains within acceptable limits.

Alternatively, if the objective is to reduce the settling time, Fig.~\ref{Figure_Urban_B_40_Contour_1p5}(b) shows that this can only be achieved at the expense of an increase in the acceleration level. In other words, no suspension configuration is able to simultaneously decrease the settling time without penalizing ride comfort. Therefore, the results indicate that, for light vehicles operating on paved roads, asymmetric damping does not emerge as an optimal design requirement, thereby explaining why symmetric suspension tuning remains effective under nominal driving conditions.

\begin{figure}[h]
\centering
\includegraphics[width=130mm]{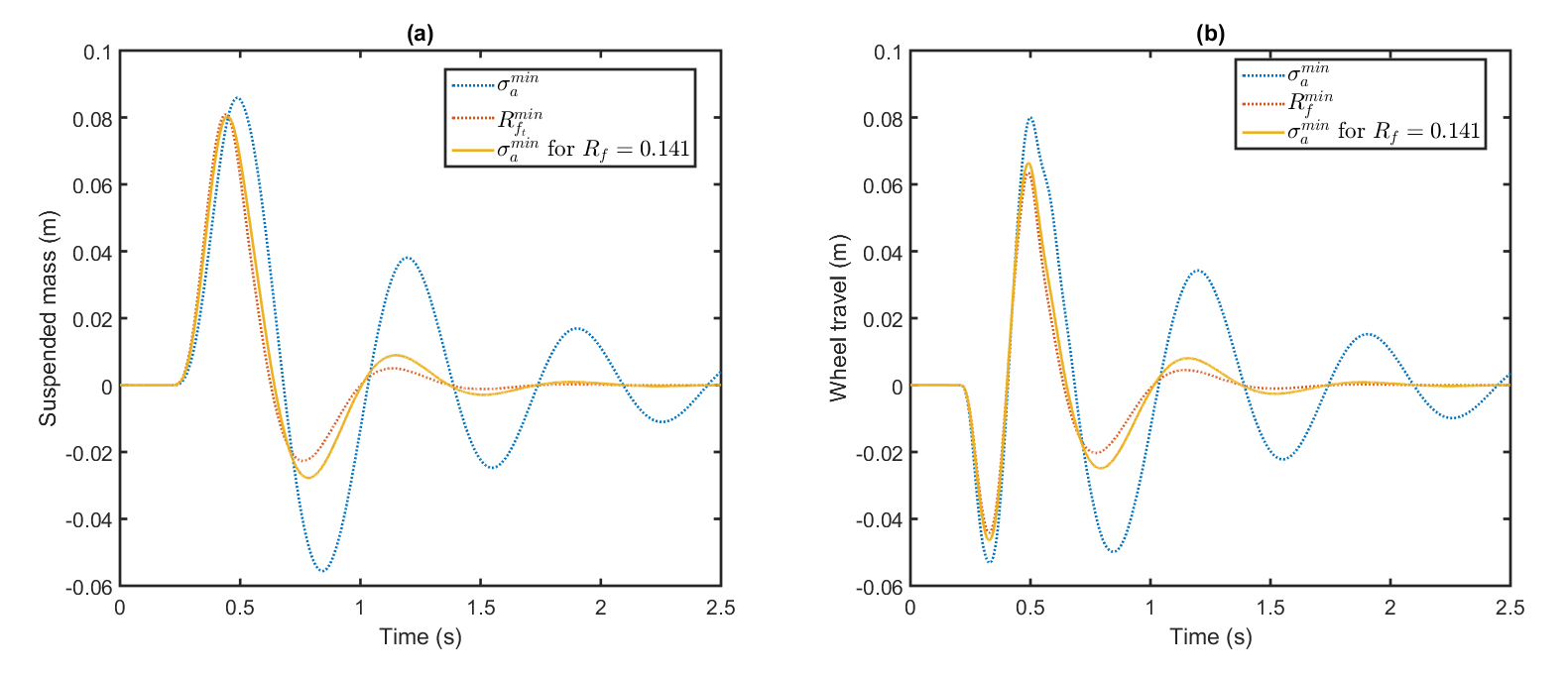}
\caption{Time histories of suspension response for a light vehicle under the bump excitation: (a) vertical displacement of the sprung mass and (b) wheel travel. In each plot, four curves are shown, corresponding to damping configurations obtained by minimizing the sprung-mass acceleration standard deviation $\sigma_{a_w}$, minimizing the tire-road contact force variation $R_{f_t}$, minimizing $\sigma_{a_w}$ subject to $R_{f_t} = 0.142$, and a reference asymmetric damping configuration.}
\label{Figure_Transient_Urban_1p5}
\end{figure}

Figure~\ref{Figure_Transient_Urban_1p5} presents the dynamic response of the same light vehicle when traversing a speed bump at a velocity of 5 m/s, considering the three suspension configurations previously identified in Fig.~\ref{Figure_Urban_B_40_Contour_1p5}. As already observed in the steady-state analysis, the configuration optimized for minimum weighted acceleration leads to a markedly long transient response, whereas the configuration optimized for minimum tire–force ratio $R_{f_t}$ allows the vehicle to recover steady motion more rapidly after the disturbance. Intermediate configurations offer a practical compromise, yielding transient responses that remain within acceptable limits while achieving lower acceleration levels at the expense of a moderate increase in $R_{f_t}$.

\begin{figure}
\centering
\includegraphics[width=130mm]{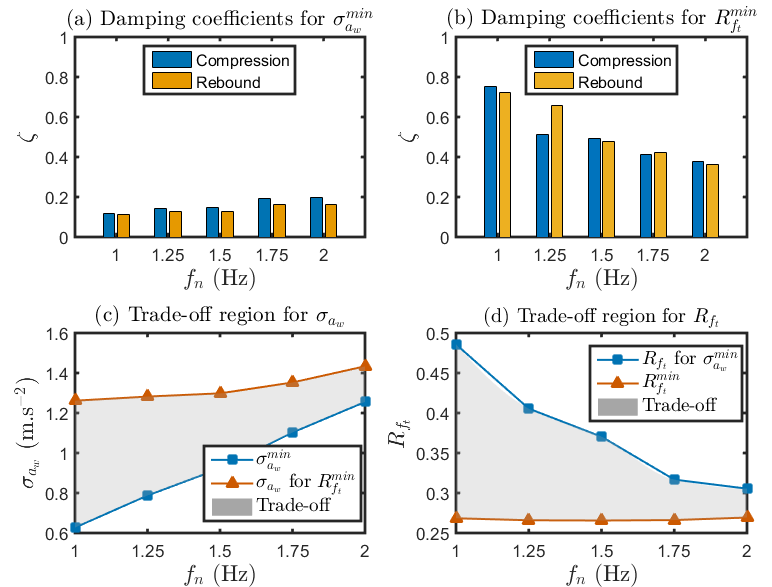}
\caption{Optimal damping coefficients for a light vehicle on a Class C road at 40 m/s, considering suspension natural frequencies of 1.0, 1.25, 1.5, 1.75, and 2.0 Hz: (a) acceleration-based optimization results; (b) tire–road contact force–based optimization results; (c) resulting acceleration levels; and (d) tire–road contact forces and corresponding trade-off regions.}
\label{Figure_Urban_C_40}
\end{figure}

The next analysis considers a light vehicle operating on an unpaved road classified as ISO Class C. Figure~\ref{Figure_Urban_C_40}(a) presents the suspension configurations obtained from acceleration-based optimization, whereas Fig.~\ref{Figure_Urban_C_40}(b) shows the corresponding configurations resulting from the minimization of $R_{f_t}$. It is immediately evident that optimization focused solely on acceleration is not feasible in this case, as it leads to values of $R_{f_t}$ well above the admissible threshold of 0.3, indicating a high likelihood of loss of tire--road contact.

\begin{figure}
\centering
\includegraphics[width=130mm]{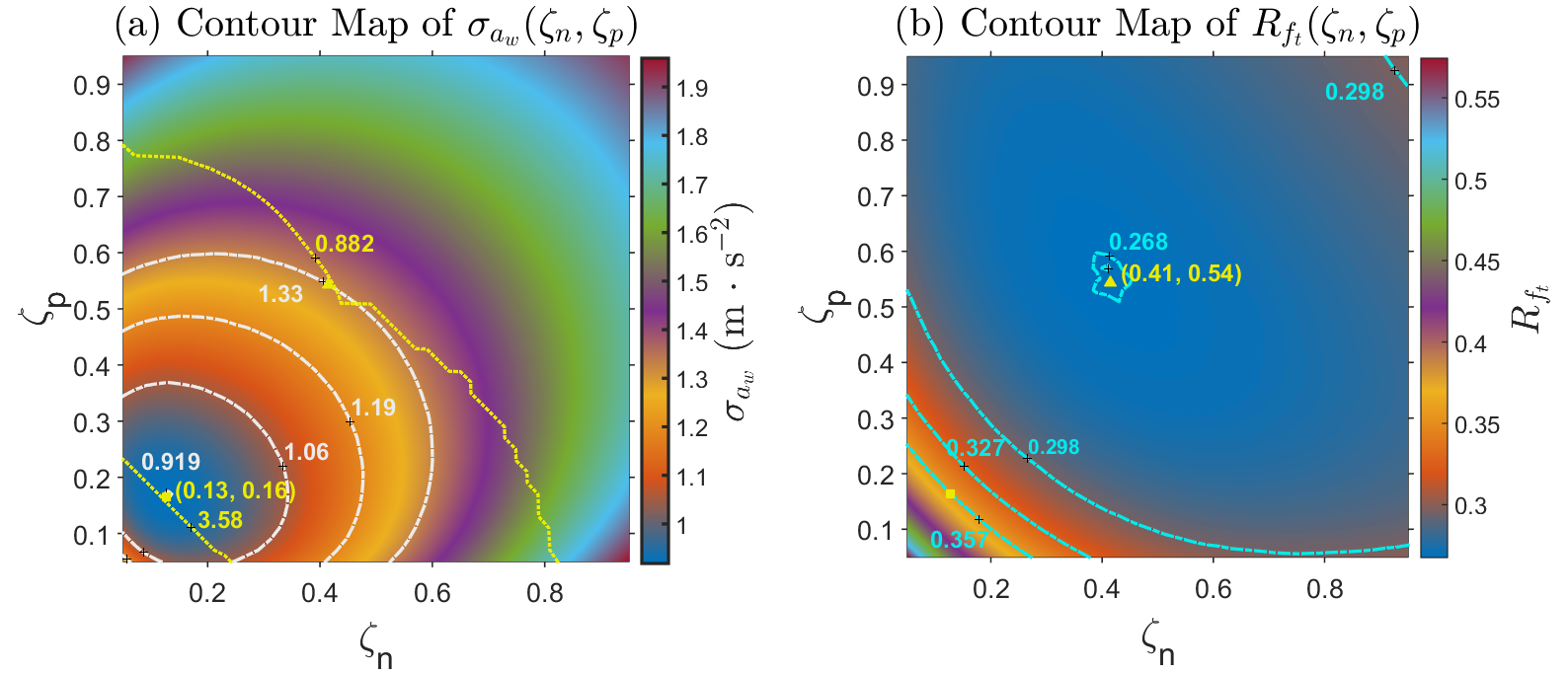}
\caption{Contour maps of suspension performance metrics for a light vehicle operating on a Class C road at a constant velocity of 40 m/s, with a suspension natural frequency of $f_n=1.5\text{ Hz}$: (a) standard deviation of the sprung-mass vertical acceleration, $\sigma_{a_w}$, used as an indicator of ride comfort; and (b) tire–road contact force variation index, $R_{f_t}$, used to assess contact stability. The contours are presented as functions of the compression and rebound damping coefficients, highlighting the influence of damping asymmetry on comfort and contact performance.}
\label{Figure_Urban_C_40_Contour_1p5}
\end{figure}

Consequently, safety considerations require that the optimization be directed toward minimizing $R_{f_t}$. This approach yields values slightly below the threshold, with a minimum of approximately $R_{f_t}=0.26$, which remains close to the upper acceptable limit. However, this improvement in contact stability comes at the expense of ride comfort. For instance, the weighted acceleration obtained for $f_n=1.5\,\text{Hz}$ is $\sigma_{a_w}=1.33\,\text{m}\cdot\text{s}^{-2}$, which can be classified as very uncomfortable according to ISO 2631, as shown in Fig.~\ref{Figure_Urban_C_40_Contour_1p5}.

This behavior contrasts with the results obtained for Class B roads, where a feasible design region exists that allows a compromise between ride comfort and contact stability. In the present case, however, the constraint on $R_{f_t}$ becomes dominant, and its minimization effectively defines the only admissible design direction to ensure safe operation.

Regarding the use of asymmetric damping, the results do not reveal any clear or systematic advantage. A comparison between Fig.~\ref{Figure_Urban_C_40} and Fig.~\ref{Figure_Urban_C_40_Contour_1p5} indicates that any apparent asymmetry in the optimal configurations is small and lacks consistency. Such deviations can be attributed to the stochastic nature of the road profile generation described by Eq.~\ref{eq_rough_road}, rather than to an inherent benefit of asymmetric damping. Therefore, no conclusive evidence can be drawn supporting a systematic performance improvement associated with damping asymmetry under these operating conditions.

\subsection{Mid-heavy vehicles}

The previous results motivate an investigation into the role of the vehicle’s total mass with respect to ride comfort and safety. As shown earlier, a light vehicle operating on a Class C road reaches a safe velocity limit of approximately 40 m/s, even then exhibiting a very low comfort level due to the high accelerations transmitted to the occupants. This behavior indicates that light vehicles may not be well suited for operation on very rough roads, particularly at higher speeds, since both comfort and safety tend to be significantly compromised.

From a dynamic perspective, vehicle mass plays an important role in the suspension response because it directly influences the system inertia and the resulting vibration levels transmitted to the occupants and to the tire–ground contact interface. In general, an increase in vehicle mass tends to reduce the acceleration levels associated with road excitation, potentially allowing a better balance between ride comfort and safety. However, the use of heavier vehicles is also associated with relevant drawbacks, such as the requirement for more powerful engines, increased fuel consumption and emissions, and greater road wear. Therefore, it is important to analyze how the suspension parameters should be adjusted when the vehicle mass changes.

\begin{figure}
\centering
\includegraphics[width=130mm]{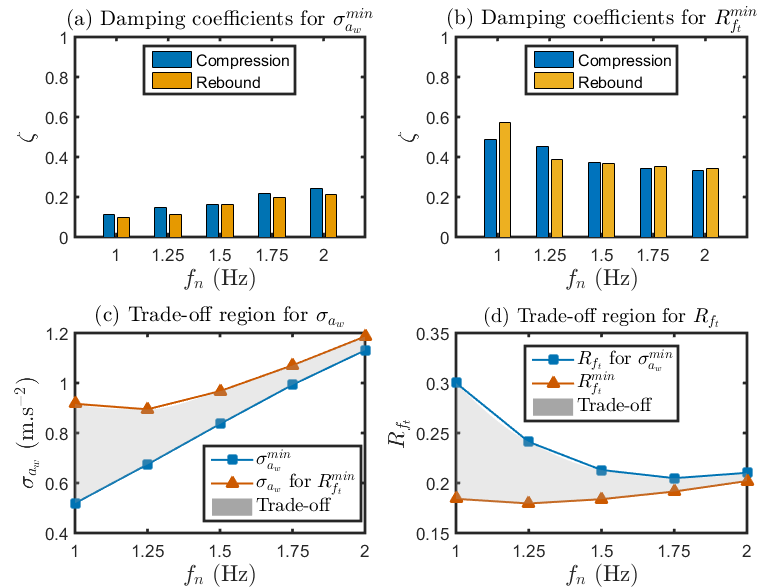}
\caption{Optimal damping coefficients for a mid-heavy vehicle on a Class C road at 40 m/s, considering suspension natural frequencies of 1.0, 1.25, 1.5, 1.75, and 2.0 Hz: (a) acceleration-based optimization results; (b) tire–road contact force–based optimization results; (c) resulting acceleration levels; and (d) tire–road contact forces and corresponding trade-off regions.}
\label{Figure_Offroad_C_40}
\end{figure}

To investigate this effect, the analysis is repeated under the same operating conditions using a heavier vehicle, allowing the relationship between vehicle mass and the optimal damping characteristics to be examined. The corresponding results are presented in Fig.~\ref{Figure_Offroad_C_40}, where a significant reduction in both $\sigma_{a_w}$ and $R_{f_t}$ is observed, indicating an improvement in both comfort and safety for this terrain condition.

\begin{figure}
\centering
\includegraphics[width=130mm]{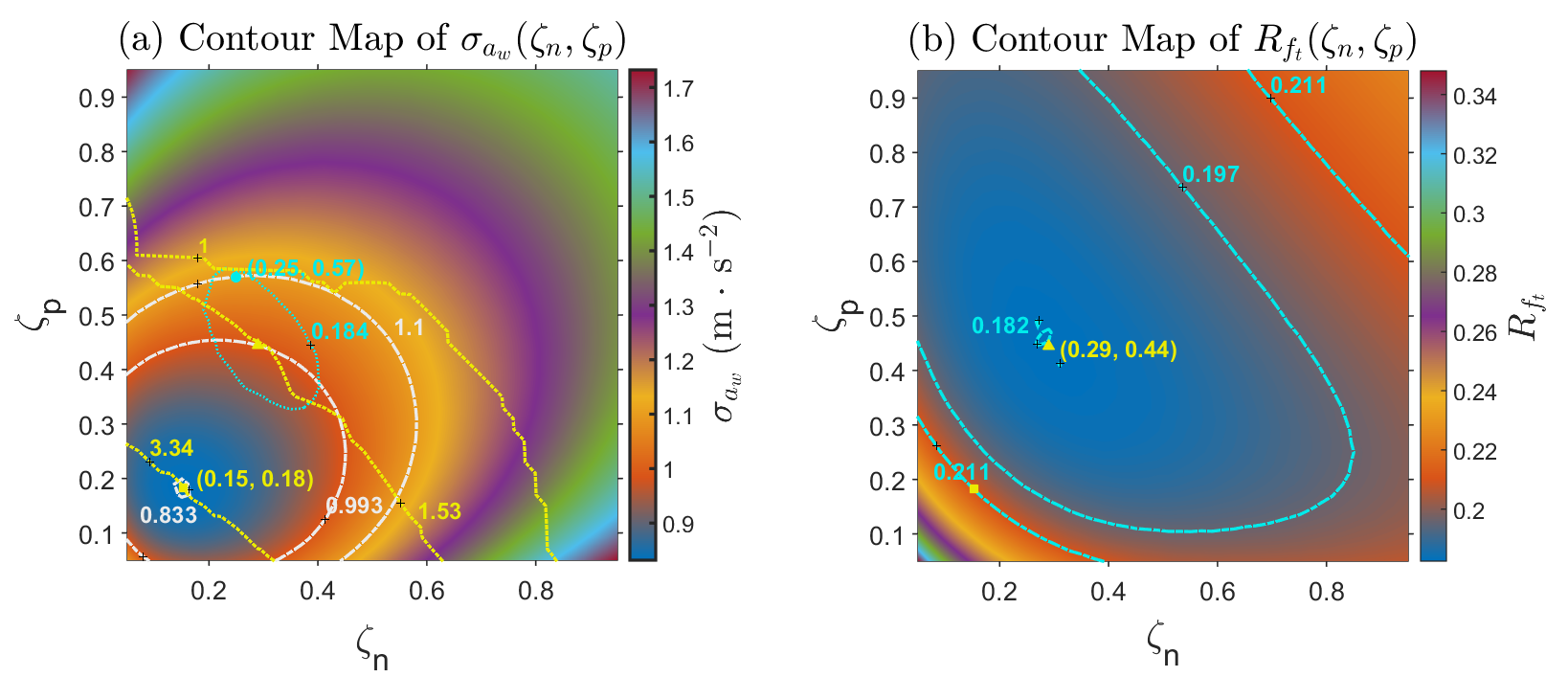}
\caption{Contour maps of suspension performance metrics for a mid-heavy vehicle operating on a Class C road at a constant velocity of 40 m/s, with a suspension natural frequency of $f_n=1.5\text{ Hz}$: (a) standard deviation of the sprung-mass vertical acceleration, $\sigma_{a_w}$, used as an indicator of ride comfort; and (b) tire–road contact force variation index, $R_{f_t}$, used to assess contact stability. The contours are presented as functions of the compression and rebound damping coefficients, highlighting the influence of damping asymmetry on comfort and contact performance.}
\label{Figure_Offroad_C_40_Contour_1p5}
\end{figure}

The contour map shown in Figure~\ref{Figure_Offroad_C_40_Contour_1p5} presents an asymmetric damper as the optimal configuration for minimizing $R_{f_t}$, contradicting the results shown in Fig.~\ref{Figure_Offroad_C_40}, for the same reason explained for the previous situation. But is is worth of note that the settling time for this adjustmente is relatively high, exceeding $t_s = 1.5$ s. If this value is considered too high, suspension configurations obtained by enforcing mutual trade-offs between $\sigma_{a_w}$ and $R_{f_t}$ — that is, configurations selected between the two limiting optimal solutions corresponding to the minimum $\sigma_{a_w}$ and the minimum $R_{f_t}$ — do not appear to be particularly advisable in this case, since these intermediate solutions tend to yield settling times, which compromises transient performance. In this context, reducing the settling time requires shifting toward more asymmetric damping configurations, which lie outside the regions associated with the minimum values of both $\sigma_{a_w}$ and $R_{f_t}$.

A compromise solution can be identified at the point $(\zeta_n,\zeta_p) = (0.26,\,0.57)$, which provides a more balanced performance. At this configuration, the settling time is reduced to approximately $t_s = 1$ s, while maintaining the acceleration level at about $1.1\ \text{m/s}^2$. Furthermore, this point still yields a near-minimum value of $R_{f_t}$, ensuring adequate tire--road contact. This result highlights the importance of considering additional dynamic performance criteria, such as settling time, in the suspension design process, as the optimal solution based solely on $R_{f_t}$ may not provide the best overall dynamic behavior.

It is worth noting that this trade-off was observed specifically for the mid-heavy vehicle operating under severe conditions (ISO Class C road at $40$ m/s) considered in this study, and may not necessarily generalize to milder operating conditions or different vehicle classes.

\begin{figure}
\centering
\includegraphics[width=130mm]{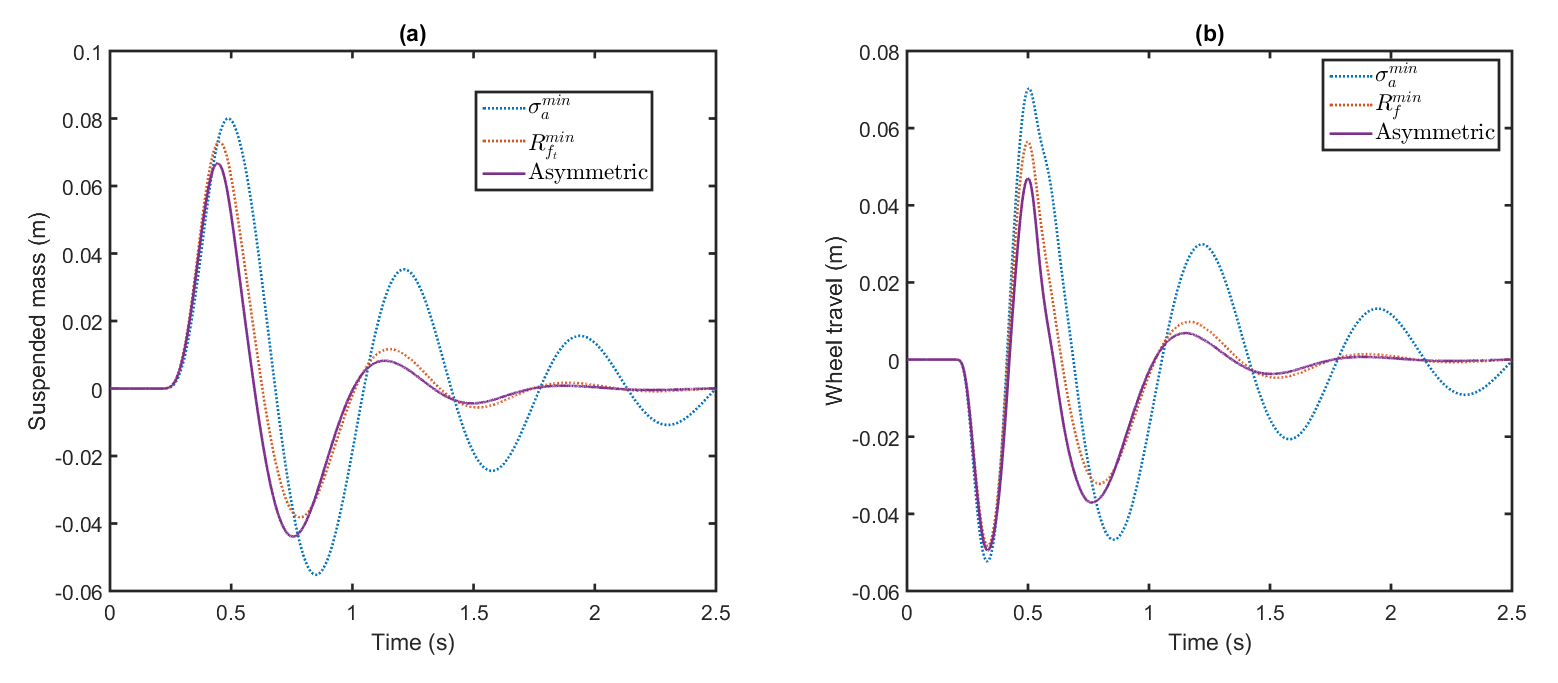}
\caption{Time histories of suspension response for a mid-heavy vehicle under the bump excitation: (a) vertical displacement of the sprung mass and (b) wheel travel. In each plot, three curves are shown, corresponding to damping configurations obtained by minimizing the sprung-mass acceleration standard deviation $\sigma_{a_w}$, minimizing the tire-road contact force variation $R_{f_t}$, and a reference asymmetric damping configuration.}
\label{Figure_Transient_Offroad_1p5}
\end{figure}

Figure~\ref{Figure_Transient_Offroad_1p5} illustrates the dynamic response of the same mid-heavy vehicle as it traverses a speed bump at a velocity of 5 m/s, considering the three suspension configurations previously identified in Fig.~\ref{Figure_Offroad_C_40_Contour_1p5}. Under this transient excitation, increasing the degree of damping asymmetry leads to a noticeable reduction in settling time and rebound wheel travel, but at the expense of reduced ride comfort and contact safety. This behavior highlights the potential of asymmetric damping to improve transient performance; however, such benefits are accompanied by a non-negligible trade-off in comfort and safety, particularly under off-road operating conditions.

\begin{figure}
\centering
\includegraphics[width=130mm]{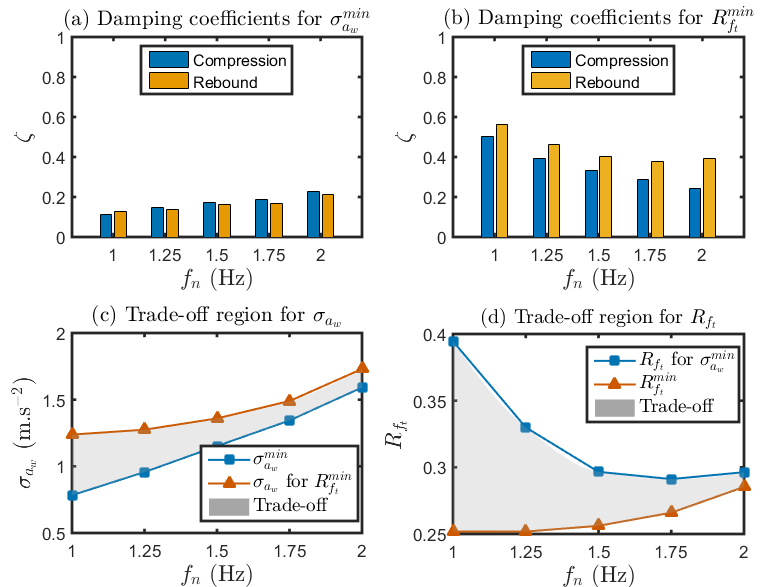}
\caption{Optimal damping coefficients for a mid-heavy vehicle on a Class D road at 20 m/s, considering suspension natural frequencies of 1.0, 1.25, 1.5, 1.75, and 2.0 Hz: (a) acceleration-based optimization results; (b) tire–road contact force–based optimization results; (c) resulting acceleration levels; and (d) tire–road contact forces and corresponding trade-off regions.}
\label{Figure_Offroad_D_20}
\end{figure}

Finally, the case of a medium–heavy vehicle operating on an ISO Class D road is analyzed. The results presented in Fig.~\ref{Figure_Offroad_D_20} indicate that, even at a relatively low speed of $20$ m/s, the value of $R_{f_t}$ approaches the limiting range of $0.25$–$0.30$. This behavior highlights the severity of the excitation associated with highly rough road profiles, for which maintaining continuous tire–ground contact becomes particularly challenging.

\begin{figure}
\centering
\includegraphics[width=130mm]{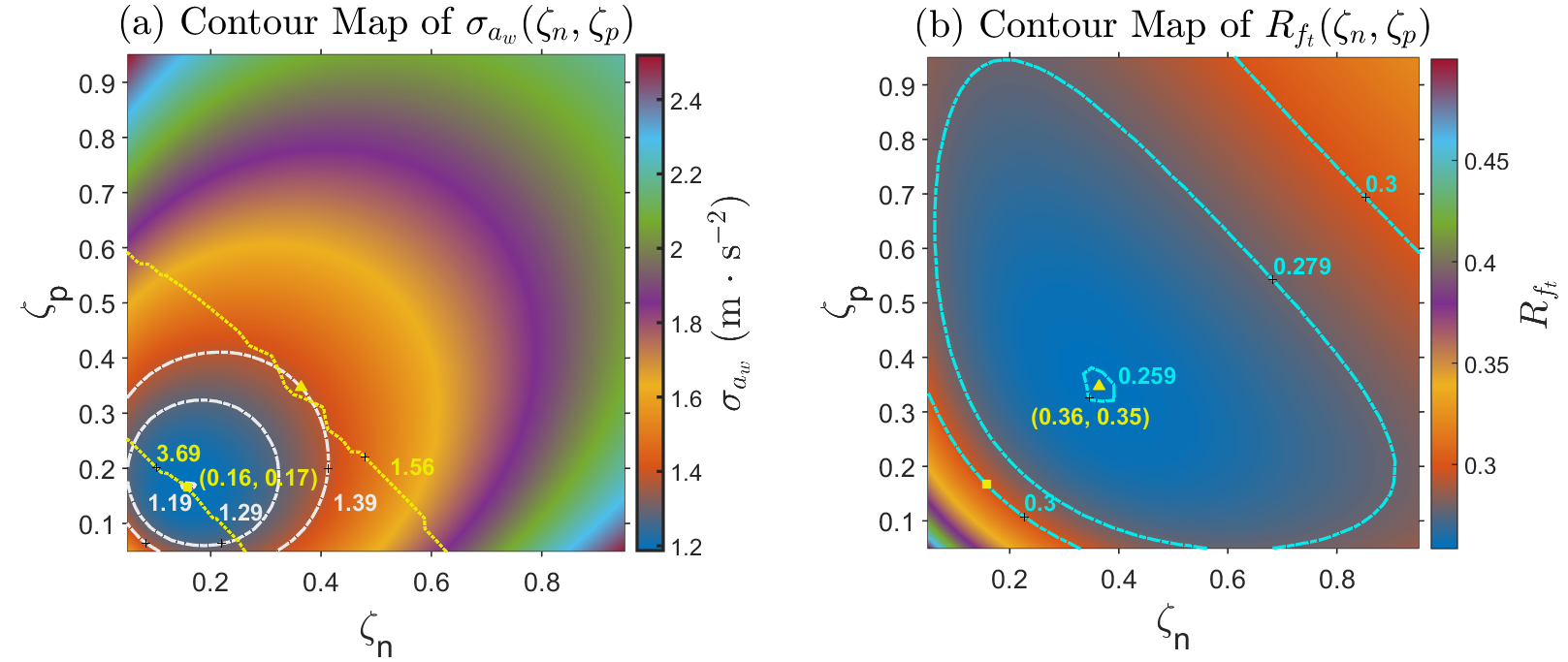}
\caption{Contour maps of suspension performance metrics for a mid-heavy vehicle operating on a Class D road at a constant velocity of 20 m/s, with a suspension natural frequency of $f_n=1.5\text{ Hz}$: (a) standard deviation of the sprung-mass vertical acceleration, $\sigma_{a_w}$, used as an indicator of ride comfort; and (b) tire–road contact force variation index, $R_{f_t}$, used to assess contact stability. The contours are presented as functions of the compression and rebound damping coefficients, highlighting the influence of damping asymmetry on comfort and contact performance.}
\label{Figure_Offroad_D_20_Contour_1p5}
\end{figure}

It is further observed that all optimal solutions point toward the use of asymmetric suspension configurations, as illustrated in Fig.~\ref{Figure_Offroad_D_20_Contour_1p5}. Similarly to the case of a light vehicle operating on an ISO Class C road at $40$ m/s, the results show that minimizing $R_{f_t}$ becomes the only feasible design strategy to preserve road-holding under such conditions.

However, this improvement in safety is achieved at the expense of ride comfort, as evidenced by the higher acceleration levels and increased settling times. This outcome clearly illustrates the inherent trade-off between comfort and safety in severe operating conditions, where prioritizing tire–ground contact imposes stricter demands on the suspension system.

Taken together, these results demonstrate that asymmetric damping does not represent a universally optimal suspension feature, but rather an emergent solution that becomes necessary as operating conditions approach critical limits. For light vehicles on smooth roads, symmetric damping provides satisfactory comfort and stability, explaining the success of traditional tuning practices. As road roughness increases or vehicle mass grows, however, optimization consistently favors configurations with higher rebound damping than compression damping. This asymmetry acts as a stabilizing mechanism that limits wheel unloading, accelerates transient decay, and preserves tire–ground contact under severe excitation.

From this perspective, commonly adopted empirical design rules prescribing stronger rebound damping can be interpreted as practical approximations of optimal solutions arising from complex, nonlinear vehicle dynamics—solutions that are now made explicit through the optimization framework proposed in this work.

\section{Conclusions} 
\label{sec5}

This work proposed a scenario-driven optimization framework for passive vehicle suspensions with asymmetric damping, with the specific goal of providing a mathematically principled explanation for empirical tuning rules widely used in practice. A minimal yet representative quarter-car model was adopted to isolate the key mechanisms governing the comfort--safety trade-off, while standardized stochastic road excitations (ISO 8608) were used to connect the analysis to the road classes implicitly assumed in engineering design practice. Within this controlled abstraction, suspension performance was quantified through complementary metrics: the ISO 2631 weighted RMS acceleration $\sigma_{a_w}$ (ride comfort), the tire--ground contact force ratio $R_{f_t}$ (road holding), and the settling time (transient performance). A Cross-Entropy optimization strategy enabled systematic exploration of these competing objectives across different road roughness levels, vehicle masses, and operating velocities, using a scalable parametrization in terms of natural frequency and damping ratios.

The results demonstrate that suspension optimality is strongly scenario-dependent. For moderate road roughness and non-critical operating conditions, symmetric damping often remains near-optimal, explaining why conventional symmetric tuning can be effective in everyday driving. As excitation severity increases and the system approaches critical regimes, however, asymmetric damping emerges as a necessary design feature to maintain $R_{f_t}$ within safe limits and prevent loss of tire--ground contact. In these regimes, solutions with rebound damping substantially larger than compression damping are consistently selected, and ratios in the range of two to three (often cited as empirical practice) arise naturally as near-optimal configurations. Beyond improving road holding, increased rebound-to-compression damping was also shown to reduce settling time and improve transient behavior without a disproportionate penalty in $\sigma_{a_w}$, clarifying the practical effectiveness of asymmetric damping as a passive design mechanism.

Overall, the proposed framework bridges empirical suspension design practice and its underlying nonlinear dynamics by making explicit when damping asymmetry is unnecessary, when it becomes mandatory, and which asymmetry levels are justified by the operating scenario. Future work may extend this principled approach to richer vehicle models and additional effects (e.g., nonlinear springs, damper hysteresis, and pitch/roll coupling) while preserving the central benefit of scenario-dependent, optimization-based interpretability.

\paragraph{Funding}

This work was supported by the National Council for Scientific and Technological Development (CNPq) under Grant No.~305476/2022-0; the Coordination for the Improvement of Higher Education Personnel (CAPES), Finance Code~001; and the Carlos Chagas Filho Research Foundation of the State of Rio de Janeiro (FAPERJ) through Grants 211.037/2019 and 204.477/2024.

\paragraph{Code Availability}
The MATLAB code developed for this study is available in the OptSusp repository: \url{https://github.com/americocunhajr/OptSusp} . 

\paragraph{Conflict of Interest }
The authors declare that they have no conflict of interest.

\paragraph{Disclaimer}

This manuscript benefited from grammatical review and stylistic refinement assisted by artificial intelligence–based tools, including Grammarly and ChatGPT. The authors retain full responsibility for the content, interpretation, and final wording of the text.



\end{document}